\documentclass[aps,prl,twocolumn,superscriptaddress,shortbibliography]{revtex4-1}
\usepackage{blindtext}
\usepackage{graphicx}
\usepackage{comment}
\usepackage{amsmath,amssymb}
\usepackage{xcolor}
\usepackage{dsfont}

\usepackage{appendix}
\usepackage{xparse}
\usepackage{array}
\usepackage{bm}

\usepackage{hyperref}
\usepackage{comment}

\newcommand{\ket}[1]{|{#1}\rangle}
\newcommand{\bra}[1]{\langle{#1}|}

\newcommand{\sx}{\hat{\sigma}^x}

\begin{document}

\author{Marc Nairn}
\affiliation{Institut für Theoretische Physik, Universität Tübingen, Auf der Morgenstelle 14, 72076 Tübingen, Germany}

\author{Luigi Giannelli}
\affiliation{Dipartimento di Fisica e Astronomia “Ettore Majorana”, Universit\`a di Catania, Via S. Sofia 64, 95123 Catania, Italy}
\affiliation{INFN, Sezione di Catania, 95123, Catania, Italy}

\author{Giovanna Morigi}
\affiliation{Theoretische Physik, Universität des Saarlandes, Campus E26, D-66123 Saarbrücken, Germany}

\author{Sebastian Slama}
\affiliation{Center for Quantum Science and Physikalisches Institut, Universität Tübingen, Auf der Morgenstelle 14, 72076 Tübingen, Germany}

\author{Beatriz Olmos}
\affiliation{Institut für Theoretische Physik, Universität Tübingen, Auf der Morgenstelle 14, 72076 Tübingen, Germany}

\author{Simon B. Jäger}
\affiliation{Physics Department and Research Center OPTIMAS, University of Kaiserslautern-Landau, D-67663, Kaiserslautern, Germany}

\title{Spin self-organization in an optical cavity facilitated by inhomogeneous broadening}
\begin{abstract}
   We study the onset of collective spin self-organization in a thermal ensemble of driven two-level atoms confined in an optical cavity. The atoms spontaneously form a spin-pattern above a critical driving strength that sets a threshold and is determined by the cavity parameters, the initial temperature, and the transition frequency of the atomic spin. Remarkably, we find that inhomogeneous Doppler broadening facilitates the onset of spin self-organization. In particular, the threshold is non-monotonic when increasing the spin transition frequency and reaches a minimum when the Doppler broadening is of similar magnitude. This feature emerges due to Doppler-induced resonances. Above the threshold, we find cooperative dynamics of spin, spatial, and momentum degrees of freedom leading to density modulations, fast reduction of kinetic energy, and the emergence of non-thermal states. More broadly, our work demonstrates how broadening can facilitate strong light-matter interactions in many-body systems. 
\end{abstract}

\date{\today}

\maketitle 

{\em Introduction}.-- 
The study of strong light-matter interactions is an active research field including fundamental studies on out-of-equilibrium phases~\cite{Baumann:2010,Klinder:2015,Mivehvar:2019,Mivehvar:2021,Landig:2016,Marsh:2021,kroeze:2023,Guo:2021,Schmidt:2014,Slama:2007} and exotic dynamics in open quantum systems~\cite{Iemini:2018,Gong:2018,Defenu:2024,Young:2024,Schuster:2020}.  Strong light-matter interactions can be realized by placing atoms, molecules or materials in an optical cavity with practical applications in new generations of sensors and clocks~\cite{Meiser:2010,Bohnet:2012,Ludlow:2015,Pezze:2018,Luo:2024}, cavity-enhanced chemistry~\cite{Lidzey:1998,Lidzey:2000,Schwartz:2011,Herrera:2016,Hirai:2023}, as well as metamaterials such as light-induced superconductors~\cite{Koppens:2011,Cavalleri:2018,Torre:2021}. A workhorse for describing features in such systems are the Tavis-Cummings and Dicke models~\cite{Kirton:2019,Larson:2021,Thompson:1992,Hertzog:2019}. The latter is famous for the prediction of a phase transition to a superradiant state that exhibits macroscopic coherence in the cavity field and the atomic medium above a critical light-matter coupling strength~\cite{Dimer:2007,Baumann:2010,Zhiqiang:2017}. This coupling strength is determined by the transition frequency of the cavity-coupled states, while the broadening of this transition is generally considered detrimental. In particular, for thermal gases that involve large inhomogeneous broadening the critical coupling strength is determined by the thermal energy~\cite{Black:2003,Domokos:2002,Asboth:2005,Arnold:2012,Schuetz:2016}. Similar effects are also relevant in Fermi gases~\cite{Keeling:2014,Piazza:2014,Zhang:2021,Helson:2023,zwettler:2024,Marijanovic:2024} where the Fermi energy can determine the threshold. The bridge between these two regimes from a single well-resolved transition frequency to a broad distribution of frequencies is, however, largely unexplored although relevant not only for thermal atoms but also molecules and materials that are naturally prone of inhomogeneous broadening.  

\begin{figure}[t]
    \centering
    \includegraphics[width=1\linewidth]{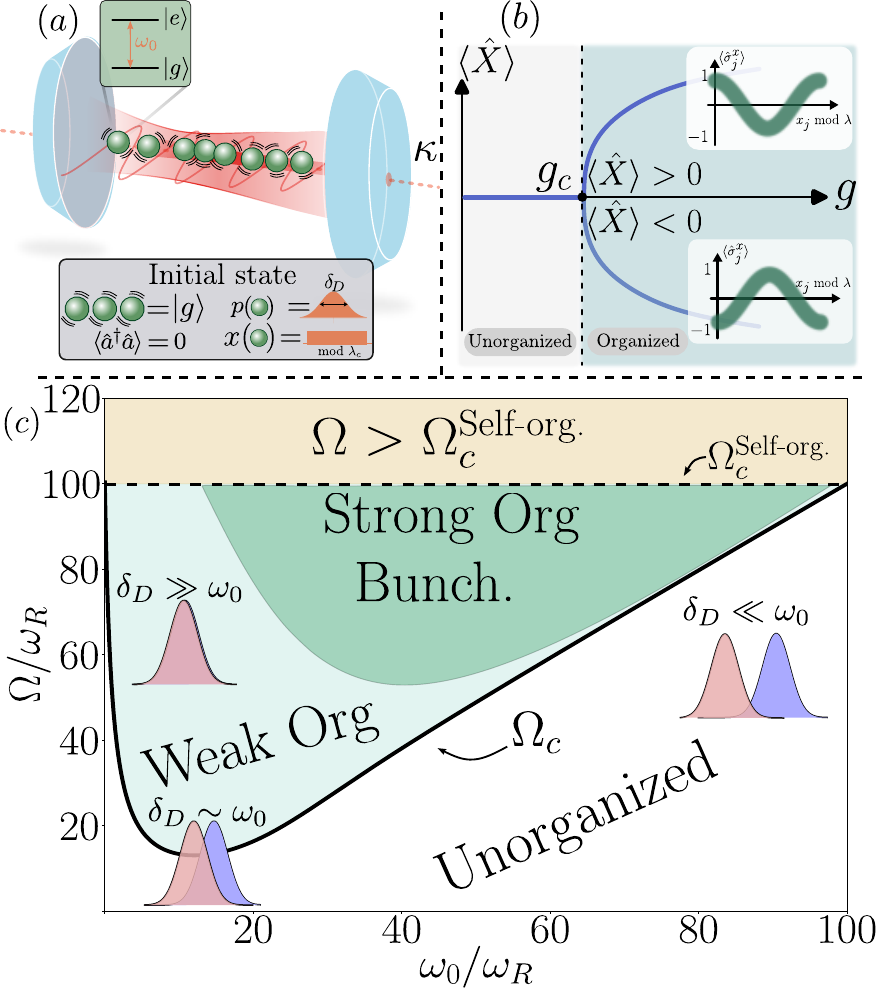}
    \caption{\textit{Cavity setup and spin self-organization.} \textbf{(a):} Schematic of the atom-cavity configuration and the initial state of the dynamics. \textbf{(b):} Emergence of the spin self-organization phase transition as spin-position correlations contained in order parameter $\langle \hat{X}\rangle$ grow once the coupling strength $g$ is tuned beyond the critical  $g_c$. The symmetry-broken $\langle \hat{X}\rangle\neq0$ phase splits into two real-space spin-aligned configurations. \textbf{(c):} Phase diagram for a thermal Doppler width $\delta_D=10\omega_R$, with the critical coupling solid line $\Omega_c$ marking the onset of (weak) spin self-organization obtained from the stability analysis \eqref{eq:critical_coupling_integral}. {The dashed line represents the expected threshold from the classical atom limit $\Omega_c^\text{Self-org.}$ (see main text). The darker green region bounds the strongly organized phase accompanied by bunching in the form of spin density gratings.}} 
    \label{fig:Fig1}
\end{figure}

{In this work, we fill this gap based on a model of thermal atoms with two ground states (spins) whose transition frequency can be tuned and is inhomogeneously broadened due to the Doppler effect. By external driving, one can realize strong coupling between the spins and the cavity field. If this coupling exceeds a threshold, we find a superradiant transition that is accompanied by spatio-temporal spin organization that we call spin self-organization. In previous works on self-organization of thermal atoms~\cite{Domokos:2001,Asboth:2005,Schuetz:2015} this threshold has been found to be proportional to the temperature of the atoms. This limit is only achieved, however, if the width of the Doppler shift distribution (Doppler width) is much larger than the transition frequency. Remarkably, when increasing the transition frequency relative to the Doppler width we find a massive reduction of the spin self-organization threshold [see Fig.~\ref{fig:Fig1}(c)]. This threshold reaches a minimum when the transition frequency is comparable to the Doppler width, which we attribute to facilitation due to Doppler resonances, leading as well to dramatic changes in the momentum distribution. %In addition, we recover the Dicke phase transition~\cite{Dimer:2007} when the Doppler width is well below the transition frequency% but do not require that the thermal energy is below the transition energy.
We study the dynamics in this previously unexplored regime %where the threshold is well below the one previously explored in thermal gases. 
and, in particular, investigate the emergent patterns using an approach based on the truncated Wigner approximation~\cite{Schachenmayer:2015b, wootters1987wigner,Milburn:2008, polkovnikov2010phase}. We find different dynamical spin self-organization regimes: a weak one, where patterns survive only on a very short timescale, and a strong one, where the spin pattern is stable on long timescales. This work provides a theoretical description of thresholds and spin pattern formation dynamics in strongly interacting light-matter systems facilitated by moderate inhomogeneous broadening. }

{\em Physical setup}.-- The system under consideration consists of a gas of $N$ atoms at finite temperature inside a lossy cavity. The atoms are tightly trapped along the transverse directions, such that their motion is restricted to the longitudinal axis of the cavity [see Fig.~\ref{fig:Fig1}(a)]. Considering each atom as an external laser- and cavity-driven few-level system in a double Lambda configuration \cite{Dimer:2007,Baden:2014,Zhang:2018}, the higher-lying levels can be adiabatically eliminated, such that only two levels coupled via the cavity remain, $\ket{g}$ (ground state) and $\ket{e}$ (excited state), reducing each atom to a spin-$1/2$ particle. An appropriate choice of the laser parameters [see Supplementary Material (SM)~\cite{supplemental} and references therein] leads to dynamics described by the Lindblad equation 
\begin{equation}
    \partial_t \hat{\rho} = -\frac{\mathrm{i}}{\hbar}\left[\hat{H}, \hat{\rho}\right] + 2\kappa\left(\hat{a}\hat{\rho} \hat{a}^\dagger-\frac{1}{2}\{\hat{a}^\dagger \hat{a}, \hat{\rho}\}\right),\label{eq:master}
\end{equation}
where $\kappa$ and $\hat{a}$ are the decay rate and the mode annihilation operator of the cavity field. The Hamiltonian for the internal (spin and cavity photons) and external (motional) degrees of freedom reads
\begin{equation}
\begin{split}\label{eq:Hamiltonian}
    \hat{H}=&-\hbar\Delta \hat{a}^\dagger \hat{a} + \hat{H}_{\mathrm{at}}+ \hbar Ng\hat{X}\left( \hat{a}^\dagger +\hat{a}\right),
\end{split}
\end{equation}
where the cavity {field detuning with respect to the incident lasers} is given by $\Delta$. The Hamiltonian $\hat{H}_{\mathrm{at}}=\sum_j\left[\hat{p}_j^2/(2m)+\hbar\omega_0\hat{\sigma}_j^z/2\right]$ consists of the atomic kinetic and internal energies, where $\omega_0$ is the atomic inversion frequency and $\hat{\sigma}_j^z=\hat{\sigma}_j^+\hat{\sigma}_j^--\hat{\sigma}_j^{-}\hat{\sigma}_j^{+}$, with $\hat{\sigma}_j^-=\ket{g}_j\bra{e}$, $\hat{\sigma}_j^+=\ket{e}_j\bra{g}$. The cavity field is coupled collectively with strength $Ng$ to the operator
\begin{equation}
\hat{X}=\frac{1}{N}\sum_{j=1}^N \hat{\sigma}^x_j \cos{(k \hat{x}_j)},\label{eq:X}
\end{equation}
whose expectation value is the \emph{spin self-organization} order parameter. Importantly, it includes, besides the internal spin coherences $\hat{\sigma}_j^x=\hat{\sigma}_j^++\hat{\sigma}_j^-$, the information of the cavity mode function with wave vector $k$. Note that by ignoring $\cos(k\hat{x}_j)$ or $\hat{\sigma}_j^x$ we recover the Dicke~\cite{Baumann:2010, Kirton:2019, Klinder:2015, hepp:1973} or spatial self-organization models~\cite{Domokos:2002,Asboth:2005, Niedenzu:2011,Arnold:2012}, respectively. In our work, however, it is crucial that spin and motional dynamics are treated on equal footing. The build-up of $\langle\hat{X}\rangle$ is accompanied by emergent spin-position correlations [see Fig.~\ref{fig:Fig1}(b)] and results in superradiant emission into the cavity.

\textit{Stability analysis and spin organization}.-- 
We start by analyzing the dynamics of the system at very short timescales in order to identify the parameter regime where the system will undergo spin self-organization. For this, we assume that initially the cavity is empty, i.e. $\langle\hat{a}^\dagger \hat{a}\rangle=0$, the atoms are in the ground state $\ket{g}$ and in a disordered state, with the initial position of the $j$-th spin $x_j$ distributed uniformly along the cavity axis. The initial momentum distribution is thermal with (Doppler) width $\delta_D=k\sqrt{\langle\hat{p}^2\rangle}/m= k/\sqrt{m\beta}$ where $\beta=1/(k_BT)$ is the inverse temperature of the atoms [see Fig. \ref{fig:Fig1}(a)].

We perform a linear stability analysis of this unorganized atomic configuration, which is reported in the SM~\cite{supplemental}. We find that spin self-organization requires the coupling strength to overcome a threshold $g_c$ [cf. Fig.~\ref{fig:Fig1}(b)]. Defining a rescaled coupling strength as $\Omega=-2Ng^2\Delta/(\kappa^2+\Delta^2)$, this threshold is given by 
\begin{align}
    \Omega_c=\left[\int_{0}^{\infty}dt \,e^{-(\delta_Dt)^2/2}\sin([\omega_0+\omega_R]t)\right]^{-1}\label{eq:critical_coupling_integral},
\end{align}
where $\omega_R=\hbar k^2/(2m)$ is the recoil frequency. At very low temperatures, i.e. $\delta_D\ll\omega_R$, we recover previously known results from \textit{spinor self-ordering}~\cite{Mivehvar:2017,Kroeze:2018,Kroeze:2019,Chiacchio:2019}, where $\Omega_c^{\text{Spinor}}\approx \omega_0+\omega_R$: the sum of the usual Dicke threshold~\cite{Kirton:2019} and self-organization at zero temperature~\cite{Nagy:2008} determined by $\omega_0$ and $\omega_R$, respectively. The focus of this work is, however, on the regime $\delta_D\gg\omega_R$ which is fundamentally different: the threshold in spinor self-ordering is determined by the gap of only two states~\cite{Mivehvar:2017,Kroeze:2018,Kroeze:2019,Chiacchio:2019} [e.g. $\ket{g,p=0} \leftrightarrow(\ket{e,\hbar k}+\ket{e,-\hbar k})/\sqrt{2}$], in our case, the cavity field will couple many states.

Now, we may distinguish two extreme regimes: (i) When $\omega_0\gg\delta_D$, the internal transition frequency $\omega_0$ is well resolved. Here, the coupling strength needs to overcome $\omega_0$ to enable the formation of a coherent spin state and we find essentially the threshold of the Dicke model~\cite{Kirton:2019} $\Omega_c^{\text{Dicke}}=\omega_0$. (ii) When $\delta_D\gg\omega_0$, the internal structure of the atoms cannot be resolved. In this regime we find the threshold of \textit{spatial self-organization}~\cite{Asboth:2005,Niedenzu:2011,Schuetz:2015}, namely, $\Omega_c^{\text{Self-org.}}=2/(\hbar\beta)$ {proportional to the temperature. %Imitating the formula for $\Omega_c^{\text{Spinor}}$
one might now expect that for intermediate values of $\omega_0/\delta_D$ the threshold scales with the sum of $\Omega_c^{\text{Dicke}}$ and $\Omega_c^{\text{Self-org.}}$, which is bounded from below by $\Omega_c^{\text{Self-org.}}$. We find that this intuition is wrong: as shown in Fig.~\ref{fig:Fig1}(c), $\Omega_c$ is here well below the threshold set by $\Omega_c^{\text{Self-org.}}=100\omega_R$ for the value $\delta_D=10\omega_R$. %Starting from $\omega_0=0$ where $\Omega_c=\Omega_c^{\text{Self-org.}}=100\omega_R$
%In fact, we find that any finite value of $\omega_0>0$ results in a drastic reduction of $\Omega_c$ from $\Omega_c^{\text{Self-org.}}$. For the extreme case $\omega_0\gg \delta_D$, we find $\Omega_c\approx\Omega_c^{\text{Dicke}}$ but, even more remarkably, for finite but large $\omega_0$ we find $\Omega_c<\Omega_c^{\text{Dicke}}$.
This shows that small but finite values of $\delta_D\sim\omega_0$ facilitate spin self-organization. Moreover, the non-monotonic behavior of $\Omega_c$ is only present in the regime $\delta_D\gg\omega_R$. %and is observed in \textit{spinor self-ordering}.
It is a consequence of resonances that emerge when the Doppler-shift of some atoms matches the transition frequency of the spin. Consequently, we find a minimum of $\Omega_c$ reached for $\delta_D\sim\omega_0$ where a large amount of atoms have a vanishing effective transition frequency due to their Doppler shift}. In this regime we find rich dynamics, as we will discuss in the following.

\begin{figure*}[t]
    \centering
    \includegraphics[width=1.\textwidth]{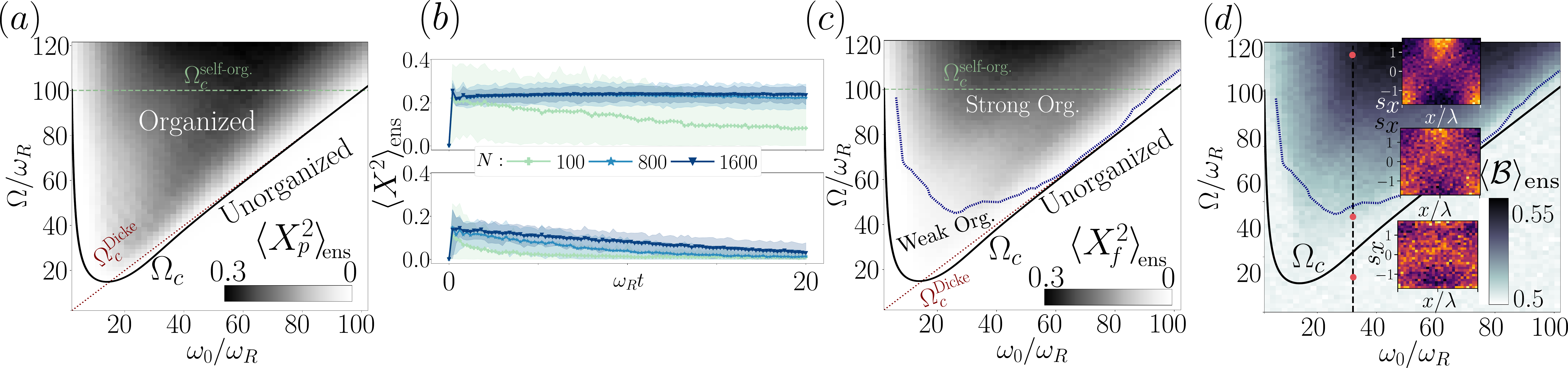}
    \caption{\textit{Spin self-organization and bunching.} Dynamics for 100 trajectories with the parameters: $\kappa = 100\omega_R,\, \Delta = -100\omega_R,$ and $\delta_D=10\omega_R$. \textbf{(a):} Peak value $\langle X_p^2\rangle_\mathrm{ens}$ of the numerical simulation for {$N=1600$} atoms. The solid black line represents the theoretically predicted threshold $\Omega_c$ from the stability analysis (\ref{eq:critical_coupling_integral}). \textbf{(b):} Dynamics for $N=100,\ 800$ and $1600$ in different regions of the phase diagram corresponding to strong spin self-organized (upper panel) and weak spin self-organized (lower panel) behaviours. The shaded regions bound 90\% of the trajectories. \textbf{(c):} Final value $\langle X_f^2\rangle_\mathrm{ens}$ at {$t_f\omega_R=20$}, showcasing that only a portion of the phase diagram can host strong spin self-organization, i.e. one that is robust in time. {The dotted line represents the crossover point between the weak and strong self-organized phases which we spot by measuring the \textit{persistance time} $\tau$ \cite{supplemental}}. \textbf{(d):} Bunching parameter $\langle\mathcal{B}\rangle_\mathrm{ens}$ at {$t_f\omega_R=20$}. Regions with $\langle\mathcal{B}\rangle_\mathrm{ens}=0.5$ correspond to a homogeneous spin density across the cavity. The emergence of an increasingly bunched pattern at the mode-function antinodes as $\Omega$ is increased is showcased in the cut at $\omega_0/\omega_R=30$, where $0.5<\langle\mathcal{B}\rangle_\mathrm{ens}\leq 1$ (only one of two possible patterns is shown here, with 1000 trajectories for improved resolution).} 
    \label{fig:Fig2}
\end{figure*}

\textit{Numerical simulation of the dynamics}.-- To study the dynamics in the regime where $\delta_D\gg\omega_R$, we combine two established methods~\cite{Giannelli:2020}: a discrete truncated Wigner simulation for the spins~\cite{Schachenmayer:2015b, wootters1987wigner}, and truncated Wigner simulation for the semiclassical dynamics of continuous variable systems~\cite{polkovnikov2010phase, Milburn:2008}. We simulate trajectories of the semiclassical fields $a_x\sim (\hat{a}+\hat{a}^\dagger)/2$ $a_{p}\sim\mathrm{i}(\hat{a}^\dag-\hat{a})/2$, positions $x_j$, momenta $p_j$ including opto-mechanical forces, and Pauli matrices $s_j^\alpha\sim\hat{\sigma}_j^\alpha$ with $\alpha=x,y,z$. Their dynamics are coupled by stochastic differential equations, and expectation values can be calculated by performing an ensemble average $\langle\,.\,\rangle_{\mathrm{ens}}$ over many realizations of the initial state (for equations of motion and further information see SM~\cite{supplemental}).  To describe the initial state, we set $s_j^z=-1$ and sample $s_j^\alpha=\pm1$ ($\alpha=x,y$) independently. The atomic positions are chosen randomly and uniformly from $[0,2\pi/k)$ and the momenta are sampled from a Gaussian with zero mean and variance $\langle p^2\rangle_{\mathrm{ens}}=m/\beta$. To incorporate the initial vacuum fluctuations in the cavity field, we sample $a_x$ and $a_p$ from a Gaussian with zero mean and $\langle a_x^2\rangle_{\mathrm{ens}}=1/4=\langle a_p^2\rangle_{\mathrm{ens}}$.

Using this simulation method we calculate the time evolution of the order parameter $\langle X^2(t) \rangle_{\mathrm{ens}}$ for a fixed value of  $\delta_D=10\omega_R$ varying the atomic frequency $\omega_0$ and the coupling $\Omega$. For low enough values of $\Omega$, the system remains for all times in the unorganized phase, i.e. $\langle X^2(t) \rangle_{\mathrm{ens}}\approx0$. As the coupling is increased, this unorganized phase is no longer stable and the system \emph{jumps} at very short times $t_p$ into the spin self-organized phase, confirmed by a value $\langle X^2_p \rangle_{\mathrm{ens}}\equiv \langle X^2(t_p) \rangle_{\mathrm{ens}}>0$. This timescale $t_p$ is determined by the collective coupling strength of the atoms to the cavity which is for our parameters much shorter than $1/\omega_R$ [see Fig.~\ref{fig:Fig2}(b)]. In Fig.~\ref{fig:Fig2}(a) we have plotted $\langle X^2_p \rangle_{\mathrm{ens}}$ for a range of values of $\omega_0$ and $\Omega$, and observe an almost perfect coincidence of the numerically obtained transition from spin unorganized to self-organized phase with the line predicted by our stability analysis.

Our stability analysis does not predict, though, whether the spin self-organized phase reached at short times {becomes meta-stable}. Actually, as depicted in Fig.~\ref{fig:Fig2}(b), the numerical evolution of the system for a finite amount of atoms $N$ can decay back into an unorganized state. In order to determine whether spin self-organization is present { after long times and in the thermodynamic limit, we perform simulations for increasing atom numbers [see Fig.~\ref{fig:Fig2}(b)].} We observe $\langle X^2\rangle_{\mathrm{ens}}\to 0$ for low values of $\omega_0$, as $\omega_Rt\gg 1$. Here, the internal energy of the spins is insufficient to sustain organization on longer times and dissipation channels become commensurate with the Hamiltonian dynamics, resulting in a relaxation back to the unorganized state. For large enough values of $\Omega$ and $\omega_0$, however, the spins remain organized. To highlight these different regimes we have depicted in Fig.~\ref{fig:Fig2}(c) the observable $\langle X^2_f \rangle_{\mathrm{ens}}\equiv \langle X^2(t_f) \rangle_{\mathrm{ens}}$ {at $t_f=20/\omega_R$}. The timescales $t_f$ is chosen such that the spins have sufficient time to redistribute themselves over several wavelengths.
We refer to spin self-organization on these very different timescales, $t_p$ and $t_f$, as weak and strong spin self-organized phases, respectively. {We define the transition between them formally by introducing the \textit{persistance time} $\tau$ after which the $\langle X^2\rangle_{\mathrm{ens}}$ has decayed back to half of its peak value. In the strong self-organization value we find $\tau/t_f=1$, while in the weak self-organization regime we find $\tau/t_f<1$. The calculated transition line between the two is visible as a blue dashed line in Fig.~\ref{fig:Fig2}(c). For more details, we refer the reader to the SM~\cite{supplemental}.}

{Let us emphasize that both weak and strong spin self-organization can occur well below the usual threshold $\Omega_c^{\text{Self-org.}}=100\omega_R$. In this regime,} the initial temperature is too high to allow for conventional atomic self-organization into a spatial density pattern. From this point of view the light-forces are insufficient to trap the atoms at anti-nodes of the cavity mode. This insight together with $\langle X^2_f \rangle_{\mathrm{ens}}>0$ suggests the spin exhibits a position dependent alignment $s_j^x\propto\cos(kx_j)$ and the density is homogeneous in space. Remarkably and against this intuition, we find that the density also exhibits a spatial pattern. To demonstrate this ordering, we calculate the bunching parameter $\mathcal{B}=\sum_{j=1}^N\cos^2(kx_j)/N$ \cite{Ritsch:2013} at time $t_f$ and show the result in Fig.~\ref{fig:Fig2}(d), where a density grating is characterized by values $\langle\mathcal{B}\rangle_{\mathrm{ens}}>1/2$. We find that the strong spin self-organized phase is indeed accompanied by a density modulation. The resulting state is one where both cavity-modulated spin-alignment and real-space density waves are present, as can be observed in the insets of Fig.~\ref{fig:Fig2}(d). 
Our findings show that density gratings and spin self-organization are possible for temperatures that are \textit{a priori} too high for ordinary self-organization. To understand this behavior further, we go on to study the kinetic energy and atomic momentum distribution.

\begin{comment}
For example, if we wish to evaluate the time evolution of $\langle\sx_j\rangle$ for atom $j$, we prepare $n_T$ initial conditions for the symmetrically ordered classical spin variable $s_j^x$. For each trajectory $i$ there are only two possible initial states $\alpha$ to choose from in the discrete phase-space  with $w^{\alpha}=1/2$ probability for each:

\begin{align}
\langle \sx_j\rangle (0) &= \sum_{\alpha} w^\alpha s_{j,i}^x(0)= \langle s^x_j \rangle (0) = 0\quad \text{with } s^x_{j,i}(0) = \pm1\\
    &\langle\sx_j\rangle (t) \approx \frac{1}{n_T}\sum_i^{n_T} s^x_{j,i}(t) = \langle s^x_j \rangle (t) 
\end{align}

where the time evolution for $s_{j,i}^x$ is given by the c-number Langevin equations outlined in \ref{eq:cHL1}-\ref{eq:cHLend}.
\end{comment}

\begin{figure}[t]
    \centering
    \includegraphics[width=1.\linewidth]{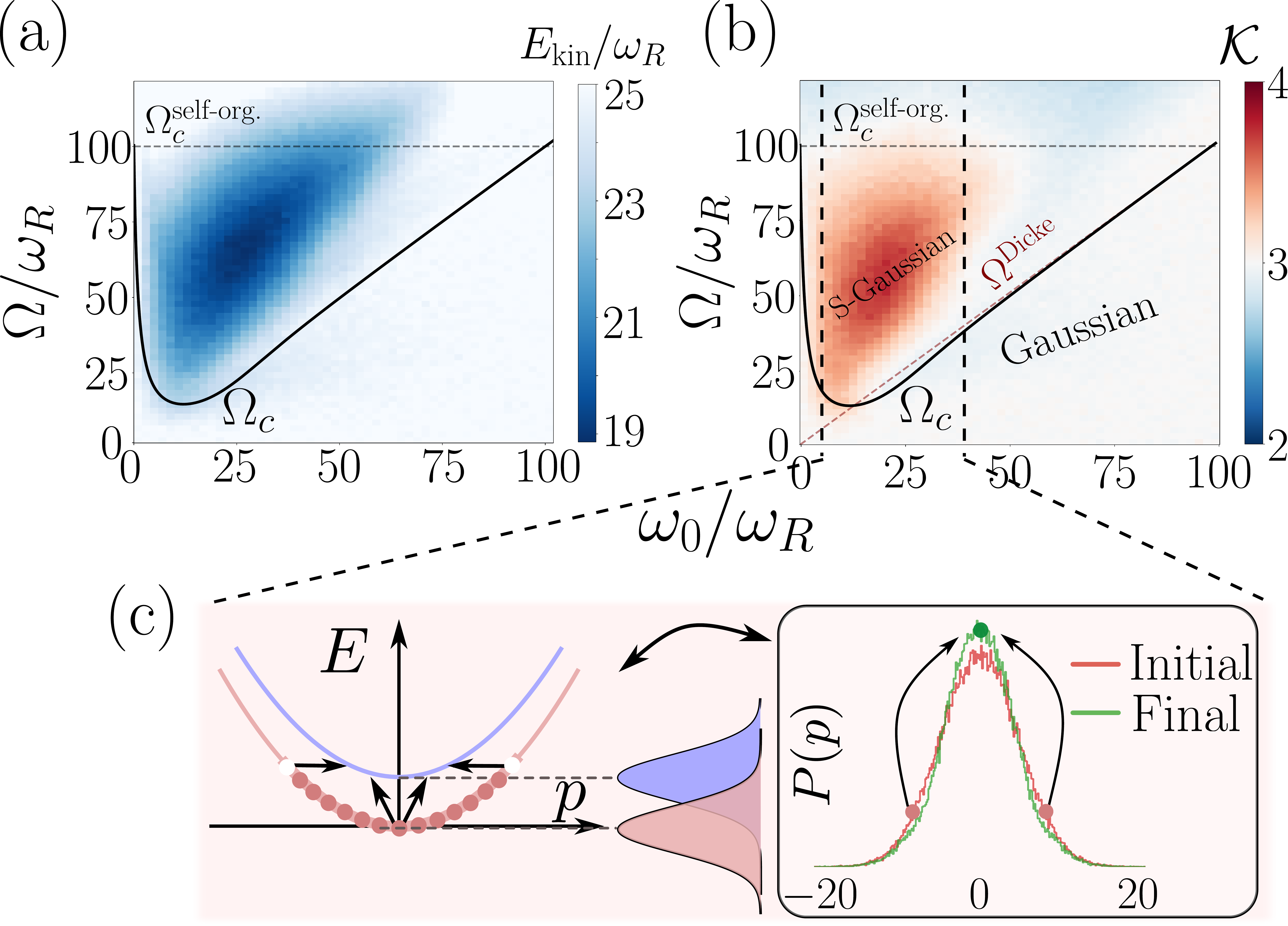}
    \caption{\textit{Final momentum state statistics.} \textbf{(a):} Kinetic energy at time $t_f=20/\omega_R$ in units of $\hbar\omega_R$, for an initial kinetic energy of $E_\text{kin} = 25\hbar\omega_R$ ($\delta_D=10\omega_R$). Spin self-organization is clearly accompanied by cooling. \textbf{(b):} Kurtosis $\mathcal{K}$ of the final momentum distribution. Regions with $\mathcal{K}>3$, which have tails that asymptotically approach zero more slowly than a Gaussian, are clearly identified in the transition region $\Omega=\omega_0\gg\delta_D$ and where $\Omega>\omega_0\sim\delta_D$. \textbf{(c):} Sketches of the energy-momentum distribution and favoured transitions (black arrows) for a fixed momentum distribution of width $\delta_D$ at the non-gaussian region $\omega_0\sim\delta_D$. Transitions from the ground to the excited state at the tails of the momentum distribution are strongly favoured, leading to an effective reduction of the energy gap the system needs to overcome to spin self-organize.} %At the transition region, $\Omega\sim\omega_0$ (right panel), these transitions are simply the only ones energetically allowed.}
    \label{fig:fig3}
\end{figure}

\textit{Momentum characteristics}.-- In Fig.~\ref{fig:fig3}(a) we show the mean kinetic energy $E_{\mathrm{kin}}=\langle\sum_jp_j^2/(2m)\rangle_{\mathrm{ens}}/N$ after the time $t_f$. We find a significant decrease of the kinetic energy with respect to the initial value $E_{\mathrm{kin}}(t=0)=25\hbar\omega_R$ in the spin self-organized region. %Dynamically, the atoms exchange energy between their internal and external degrees of freedom.
{A simple model to understand this feature can be derived by adiabatically eliminating the cavity, $\hat{a}_{\mathrm{ad}}\approx Ng\hat{X}/(\Delta+i\kappa)$, relying on $\omega_0,\delta_D\ll \Delta,\kappa$. Using  $\hat{a}=\hat{a}_{\mathrm{ad}}$ in Eq.~\eqref{eq:Hamiltonian}, we find that the atomic dynamics are described by 
\begin{equation}
\hat{H}_{\mathrm{ad}}=\hat{H}_{\mathrm{at}}-\frac{\hbar \Omega N}{2}\hat{X}^2.\label{eq:Had}
\end{equation}
}For energy to be conserved, thus, an increase of $\langle\hat{X}^2\rangle$ is always accompanied by an increase of $\langle\hat{H}_{\mathrm{at}}\rangle$. Since the atoms start in the internal ground state, their internal energy can only increase on short timescales. However, since $\hat{X}$ couples internal and external degrees of freedom, it provides also a channel for energy exchange. An atom with momentum $p>0$ can now exhibit a recoil which decreases or increases its kinetic energy. The corresponding frequency gaps are given by $\Delta E_{-}(p)\approx\omega_0-kp/m$ and $\Delta E_{+}(p)\approx \omega_0+kp/m$, where we dropped the contribution of the recoil energy. Since the process that results in reduction of kinetic energy is closer to resonance $|\Delta E_{-}|<|\Delta E_{+}|$ the atoms will more likely couple to lower momentum states. This is confirmed by our simulations, where we find that $E_{\mathrm{kin}}$ reduces on very short timescales $\sim t_p$. We remark that this process is very distinct from cavity cooling~\cite{Vuletic:2000,Horak:1997,Maunz:2004,Hosseini:2017}. The process that we describe here is a collective transfer of energy from external to internal degrees of freedom. Furthermore, it is striking that the phase diagram and some dynamical features share several similarities with the dynamics observed when the atoms are incoherently pumped and the cavity-atom coupling is described by the rotating wave approximation~\cite{Jaeger:2019,Jaeger:2020}. Differing from that model, the dynamics here analysed is determined by Doppler resonances and the counter-rotating terms turn out to be essential to trigger these dynamics. This opens the possibility to tune the response of the system by controlling the ratio of $\delta_D$ and $\omega_0$.

After the initial dynamics, the system may exhibit either weak or strong spin self-organization. {This transition is not clearly visible in the kinetic energy, which was lowered due to the Doppler resonances. However, this lowering is pronounced in the weak spin self-organization regime where $\omega_0\lesssim\delta_D$ and $\Omega<\Omega_c^{\mathrm{self-org}}$. We have verified that these are the important inequalities by studying various parameters (see SM~\cite{supplemental}). This dynamics originates from small energy gaps $|\Delta E_{-}(p)|\approx0$ that facilitate the spin self-organization, in a similar manner to Umklapp processes in a degenerate fermionic gas~\cite{Piazza:2014} [see Fig.~\ref{fig:fig3}(c) left-hand side].} The atoms with the corresponding momenta quickly reduce their momentum, which leads to a non-Gaussian momentum distribution with a pronounced central peak [see Fig.~\ref{fig:fig3}(c) right-hand side]. This is measured in Fig.~\ref{fig:fig3}(b) with the mean kurtosis $\mathcal{K} = \langle P^4\rangle_{\mathrm{ens}}/\langle P^2\rangle_{\mathrm{ens}}^2$, with $\langle P^n\rangle_{\mathrm{ens}}=\langle\sum_j p_j^n\rangle_{\mathrm{ens}}/N$. This quantity is $\mathcal{K}=3$ for a Gaussian, and the value $\mathcal {K}>3$ found in our case indicates a momentum distribution with a steeper peak. For larger $\omega_0$, {in the regime of} strong spin self-organization {we find} $\mathcal{K}\approx 3$. In this regime the system forms a density pattern, which makes the previous considerations only focusing on internal and kinetic energy invalid. {We highlight, however, that also in the strong organized regime the reduction of kinetic energy facilitates the formation of a density pattern.}

{\em Conclusions and Outlook}.-- This work bridges the gap between strong light-matter interactions in spins with a well-resolved transition frequency to those dominated by inhomogeneous broadening. We find a new intermediate regime of spin self-organization facilitated by Doppler resonances resulting in an interplay of spin and motional degrees of freedom. This interplay is also responsible for dramatic changes in the momentum statistics, such as an almost instantaneous reduction of kinetic energy of the gas.  Beyond the fundamental interest of our study, the regime that we explore is currently experimentally accessible in several groups, including the experiment of one of the authors~\cite{Suarez:2023} (see SM~\cite{supplemental} for experimental details). Future works can explore the long-time behavior of this model where we expect a pronounced role of dissipation, e.g. in the form of collective cavity cooling. Moreover, we believe that the theoretical results in this work will also be applicable to very different physical setups including free fermions with two different spin species where we expect to see a similar dynamical and threshold behavior. Another example are insulators where the cavity induces excitations of different momentum modes in the empty band. More generally, our work demonstrates how moderate inhomogeneous broadening can enhance strong light-matter interactions in many-body systems.

%%funding and acknowledgements
\acknowledgements
The authors acknowledge support by the state of Baden-Württemberg through bwHPC and the German Research Foundation (DFG) through grant no INST 40/575-1 FUGG (JUSTUS 2 cluster). The research leading to these results has received funding from the Deutsche Forschungsgemeinsschaft (DFG, German Research Foundation) under the Research Unit FOR 5413/1, Grant No. 465199066. S.B.J. acknowledges support from Research Centers of
the Deutsche Forschungsgemeinschaft (DFG): Projects
A4 and A5 in SFB/Transregio 185: OSCAR. S.S. acknowledges support from the Deutsche Forschungsgemeinschaft (DFG)
– Project-ID 422447846. 
This work was funded by the Deutsche Forschungsgemeinschaft (DFG, German Research Foundation) – Project-ID 429529648 – TRR 306 QuCoLiMa (“Quantum Cooperativity of Light and Matter’’), the DFG Forschergruppe WEAVE "Quantum many-body dynamics of matter and light in cavity QED" - Project ID 525057097, and the QuantERA project "QNet: Quantum transport, metastability, and neuromorphic applications in Quantum Networks" - Project ID 532771420. G.M. thanks Helmut Ritsch for helpful comments. LG acknowledges support from the PNRR MUR project PE0000023-NQSTI.

\bibliography{main}
%\nocite{*} %left for now for extra references
\end{document}

\title{Supplemental material: Spin-self-organization in an optical cavity facilitated by inhomogeneous broadening}

\author{Marc Nairn}
\affiliation{Institut für Theoretische Physik, Universität Tübingen, Auf der Morgenstelle 14, 72076 Tübingen, Germany}

\author{Luigi Giannelli}
\affiliation{Dipartimento di Fisica e Astronomia “Ettore Majorana”, Universit\`a di Catania, Via S. Sofia 64, 95123 Catania, Italy}
\affiliation{INFN, Sezione di Catania, 95123, Catania, Italy}

\author{Giovanna Morigi}
\affiliation{Theoretische Physik, Universität des Saarlandes, Campus E26, D-66123 Saarbrücken, Germany}

\author{Sebastian Slama}
\affiliation{Center for Quantum Science and Physikalisches Institut, Universität Tübingen, Auf der Morgenstelle 14, 72076 Tübingen, Germany}

\author{Beatriz Olmos}
\affiliation{Institut für Theoretische Physik, Universität Tübingen, Auf der Morgenstelle 14, 72076 Tübingen, Germany}

\author{Simon B. Jäger}
\affiliation{Physics Department and Research Center OPTIMAS, University of Kaiserslautern-Landau, D-67663, Kaiserslautern, Germany}

%Need more precise title

\date{\today}

\maketitle

\section{Adiabatic elimination of excited states}
In this section we describe all the necessary steps to, starting from the double-$\Lambda$ scheme suggested by Dimer et al. \cite{Dimer:2007, Zhang:2018, Baden:2014}, arrive at the effective two-level system we introduce in the main text.

We start with the single-particle Hamiltonian ($\hbar=1$): 
\begin{align}
    \hat{H}^{(1)} = H_C + \hat{H}^{(1)}_S + \hat{H}^{(1)}_I
\end{align}
where 
\begin{align}
    \hat{H}_C &= \omega_c\hat{a}^\dagger\hat{a},\\
    \hat{H}^{(1)}_S &= \frac{\hat{p}^2}{2m}+ \sum_{m=1,2,3} \omega_m|m\rangle\langle m| + \left(\Omega_1e^{-i\omega_{L1}t}|2\rangle\langle1|+ \Omega_2e^{-i\omega_{L2}t}|3\rangle\langle0|+ \text{h.c.}\right),\\
    \hat{H}^{(1)}_I & = \cos{(k\hat{x})}\left(g_1\hat{a}^\dagger|0\rangle\langle2| + g_2\hat{a}^\dagger|1\rangle\langle3| + \text{h.c.}\right).
\end{align}

The atomic energy levels are given by $\{0, \omega_1, \omega_2, \omega_3\}$, respectively, and $\omega_{L1}, \omega_{L2}$ label the pump laser frequencies driving transitions $\ket{1}\leftrightarrow\ket{2}$ and $\ket{0}\leftrightarrow\ket{3}$, respectively. The position and momentum of the atom are given by $\hat{x}, \hat{p}$ such that $[\hat{x}, \hat{p}] = i$, and $\hat{a}$ is the annihilation operator of the cavity mode with frequency $\omega_c$ and wavenumber $k$, such that $\left[\hat{a}, \hat{a}^\dagger\right] = 1$.

We move to the interaction picture to drop time-dependent terms $\propto e^{i\omega t}$ introducing the unitary $\hat{U}(t) = e^{i\hat{H}_0 t}$ with
\begin{align}
    \hat{H}_0 &= \left(\dfrac{\omega_{L1}+\omega_{L2}}{2}\right)\hat{a}^\dagger\hat{a}+\omega_{L2}|3\rangle\langle 3| + \left(\dfrac{\omega_{L2}+\omega_{L1}}{2}\right)|2\rangle\langle 2| + \left(\dfrac{\omega_{L2}-\omega_{L1}}{2}\right)|1\rangle\langle 1|.
\end{align}
The Hamiltonian in the rotating frame can then be calculated as
\begin{align}
    \hat{H}'^{(1)} &= \hat{U}\hat{H}^{(1)}\hat{U}^\dagger + i\frac{\partial \hat{U} }{\partial t}\hat{U}^\dagger \nonumber \\ 
     &=\hat{H}_C'+\hat{H}_S' + \hat{H}_I' - H_0,
    \label{eq:interaction_pict}
\end{align}
where we have defined $\hat{H}'_C=\hat{U}\hat{H}_C\hat{U}^\dag$, $\hat{H}'_S=\hat{U}\hat{H}_S\hat{U}^\dag$, and $\hat{H}'_I=\hat{U}\hat{H}_I\hat{U}^\dag$.
Summing over all these terms results in the time-independent Hamiltonian

\begin{align}
\hat{H}'^{(1)}&= -\Delta\hat{a}^\dagger\hat{a}+ \dfrac{\hat{p}^2}{2m}-\delta_1|1\rangle\langle 1|-\delta_2|2\rangle\langle2|-\delta_3|3\rangle\langle 3|\nonumber \\
&+\bigg(\Omega_1|2\rangle\langle 1| + \Omega_2|3\rangle\langle 0| + \text{h.c.}\bigg) +\cos(k\hat{x})\bigg(g_1\hat{a}^\dagger|0\rangle\langle 2|+ g_2\hat{a}^\dagger|1\rangle\langle 3| +\text{h.c.}\bigg) ,\label{eq:interaction_hamiltonian}
\end{align}

where we have introduced the detunings 
\begin{align}
&\Delta=\left(\dfrac{\omega_{L1}+\omega_{L2}}{2}\right)-\omega_c, \quad \delta_1=\left(\dfrac{\omega_{L2}-\omega_{L1}}{2}\right)-\omega_1, \nonumber\\ &\delta_2=\left(\dfrac{\omega_{L1}+\omega_{L2}}{2}\right)-\omega_2, \quad \delta_3=\omega_{L2}-\omega_3.
\end{align}
%\section{Adiabatic projector elimination}

To obtain the model that is given in the main text we need to eliminate the two excited levels $\ket{2}$ and $\ket{3}$. This is justified if the driving is strongly off-resonant so that $\delta_{2,3}\gg \delta_1, \Omega_{1,2}, g, \Delta, \kappa$. This allows us to capture the dynamics by means of an effective Hamiltonian that couples the two levels $|1\rangle\leftrightarrow|0\rangle$. Writing out the atomic part of the single-particle Hamiltonian in the four-level basis, we have
\begin{align}
    \hspace{-4mm}\hat{H}^\text{(at)}=
\begin{bmatrix}
0 & 0 & g_1\cos(k\hat{x})\hat{a}^\dagger & \Omega_2^* \\
0 & -\delta_1 & \Omega_1^* & g_2\cos(k\hat{x})\hat{a}^\dagger \\
g_1\cos(k\hat{x})\hat{a} & \Omega_1  & -\delta_2 & 0 \\
\Omega_2 & g_2\cos(k\hat{x})\hat{a}& 0 & -\delta_3
\end{bmatrix} .
\end{align}

Take the state $|\Psi\rangle = [ \alpha,   \beta,   \gamma,  \xi]^T$ satisfying the Schrödinger equation for $\hat{H}^\text{(at)}$ given by
\begin{subequations}
\begin{align}
    i\dot{\alpha}(t) &= \Omega^*_2 \xi + g_1\cos(k\hat{x})\hat{a}^\dagger \gamma \label{dynamic1}\\
    i\dot{\beta}(t) &= -\delta_1\beta + g_2\cos(k\hat{x})\hat{a}^\dagger \xi + \Omega^*_1\gamma \label{dynamic2}\\
    i\dot{\gamma}(t) &= -\delta_2\gamma + \Omega_1\beta +  g_1\cos(k\hat{x})\hat{a} \alpha \label{dynamic3}\\
    i\dot{\xi}(t) &= \Omega_2\alpha + g_2\cos(k\hat{x})\hat{a}\beta - \delta_3\xi \label{dynamic4}.
\end{align}
\end{subequations}
Since $\delta_2$ and $\delta_3$ are by far the largest frequencies in these equations we can adiabatically eliminate the dynamics of $\gamma$ and $\xi$. This is done by performing a formal integration of the dynamical equations of $\gamma$ and $\xi$, and using the result in the equations for $\alpha$ and $\beta$. This allows us to find a dynamical description within the reduced subspace spanned only by the ground states, $\ket{1}$ and $\ket{2}$, governed by 
\begin{align}
i\partial_t \begin{pmatrix} \alpha \\ \beta \end{pmatrix} &= H_\text{eff}^{(\text{at})} \begin{pmatrix} \alpha \\ \beta \end{pmatrix}
\end{align}
with effective Hamiltonian 
\begin{align}
   \hat{H}_\text{eff}^{(\text{at})} = \begin{pmatrix}
 \dfrac{g_1^2}{\delta_2}\hat{a}^\dagger\hat{a}\cos^2(k\hat{x}) +\dfrac{|\Omega_2|^2}{\delta_3} & \left(\dfrac{g_2\Omega_2}{\delta_3}\hat{a} + \dfrac{g_1\Omega_1}{\delta_2}\hat{a}^\dagger\right)\cos(k\hat{x}) \\
\left(\dfrac{g_2\Omega_2}{\delta_3}\hat{a}^\dagger + \dfrac{g_1\Omega_1}{\delta_2}\hat{a}\right)\cos(k\hat{x}) & -\delta_1 + \left( \dfrac{g_2^2}{\delta_3}\hat{a}^\dagger\hat{a}\cos^2(k\hat{x}) + \dfrac{|\Omega_1|^2}{\delta_2}\right)
\end{pmatrix}.
\end{align}
We mention that we have discarded spontaneous emission from the excited manifold, $\ket{2}$ and $\ket{3}$, in this description. This is justified if the total spontaneous emission rate $\gamma_{\mathrm{tot}}$ is much smaller than the detuning $\delta_2$ and $\delta_3$. In this case the dynamics are dominated by coherent Raman scattering processes and the incoherent scattering rate is sufficiently small.

In order to set the ground state energy for the zero cavity field at zero energy we introduce the global energy shift $\hat{U} = \exp\left(i\dfrac{|\Omega_2|^2}{\delta_3} t\right)$. Consequently, we arrive at the effective Hamiltonian 
\begin{align}
   \hat{H}_\text{eff}^{(\text{at})} &=  \omega_0 |1\rangle\langle1|+ \cos^2(k\hat{x})\bigg(U_1|0\rangle\langle0| + U_2|1\rangle\langle1|\bigg)\hat{a}^\dagger\hat{a} +g\cos(k\hat{x})\bigg(\hat{a}^\dagger|0\rangle\langle1| + \hat{a}^\dagger|1\rangle\langle0| + \text{h.c.}\bigg)
\end{align}
where we introduced $\omega_0 = \dfrac{|\Omega_1|^2}{\delta_2} - \dfrac{|\Omega_2|^2}{\delta_3} - \delta_1, U_1 = \dfrac{|g_1|^2}{\delta_2}$, $U_2 = \dfrac{|g_2|^2}{\delta_3}$ and $\dfrac{g_1\Omega_1}{\delta_2}=g=\dfrac{g_2\Omega_2}{\delta_3}$. To obtain the Hamiltonian that has been used in the main text we assumed $g_1\Omega_1/\delta_2=g_2\Omega_2/\delta_3$, we discarded $U_1$ and $U_2$ which is strictly justified if $NU_j\ll\Delta$ and $U_j\langle\hat{a}^\dag\hat{a}\rangle\ll \omega_0$. Introducing the remaining terms from \eqref{eq:interaction_hamiltonian} and applying this approach to an atomic ensemble, we obtain the Hamiltonian shown in the main text
\begin{align}
    \hat{H}&=-\hbar\Delta \hat{a}^\dagger \hat{a} + \sum_j^N\bigg[\frac{\hat{p}_j^2}{2m}+\hbar \omega_0\hat{\sigma}_j^+\hat{\sigma}_j^- + \hbar g\eta(\hat{x}_j)\left(\hat{\sigma}_j^+ +\hat{\sigma}_j^-\right)\left( \hat{a}^\dagger +\hat{a}\right)\bigg].
\end{align}
Here we defined $\hat{\sigma}_j^+=\ket{1}_j\bra{0}$ and $\hat{\sigma}_j^-=\ket{0}_j\bra{1}$ and in the main text we have slighlty changed notations $\ket{1}$ to $\ket{\uparrow}$ and $\ket{0}$ to $\ket{\downarrow}$.
\label{ap:adiabatic_elim}
\section{Derivation of critical coupling threshold}
In this section we provide details on the calculation of the critical coupling strength $\Omega_c$ above which we find spin-self-organization.

The starting point of this analysis is a quantum mean-field description of the atoms and a mean-field description of the cavity field. This assumes that the density matrix of the atoms is factorized $\hat{\rho}\approx\hat{\rho}_1\otimes\hat{\rho}_1\otimes\cdots\otimes\hat{\rho}_1$ and the cavity field is in a coherent field described by $\alpha$. Here, $\hat{\rho}_1$ is the single-particle density matrix and $\alpha$ is the coherent field amplitude. Their coupled dynamics are governed by the following equations
\begin{align}
\frac{\partial \hat{\rho}_1 }{\partial t}=&\mathcal{L}_0\hat{\rho}_1  -ig(\delta\alpha+\delta\alpha^*)[\hat{\sigma}^x\cos(k\hat{x}_j),\hat{\rho}_1],\\
\frac{d \alpha}{dt}=&- (i\Delta+\kappa)\alpha-iNg\mathrm{Tr}\left(\hat{\sigma}^x\cos(k\hat{x})\hat{\rho}_1\right),
\end{align}
where we introduced 
\begin{align}
    \mathcal{L}_0\hat{\rho}_1=\frac{1}{i\hbar}\left[\frac{\hat{p}^2}{2m}+\hbar \omega_0\hat{\sigma}^{+}\hat{\sigma}^{-},\hat{\rho}_1\right].
\end{align}

Initially we prepare the atoms in a homogeneous thermal stationary state $\hat{\rho}_0=\hat{\rho}_{\mathrm{ext}}\otimes\ket{\downarrow}\bra{\downarrow}$ with $\rho_{\mathrm{ext}}=\sum_{p}\rho_p\ket{p}\bra{p}$ which is stationary when the cavity is empty, $\alpha_0=0$.
We then study the dynamics of the fluctuations around this stationary state
\begin{align}
\hat{\rho}=&\hat{\rho}_0+\delta\rho,\\
\alpha=&\delta\alpha.
\end{align}
The dynamics of the fluctuations are governed by the following coupled differential equations
\begin{align}
\frac{\partial \delta\hat{\rho}}{\partial t}=&\mathcal{L}_0\delta\hat{\rho}-ig(\delta\alpha+\delta\alpha^*)[\hat{\sigma}^x\cos(k\hat{x}),\hat{\rho}_0] \label{eq:eom_1}\\
\frac{d\delta\alpha}{dt}=&-(i\Delta+\kappa)\delta\alpha-iNg\mathrm{Tr}\left(\hat{\sigma}^x\cos(k\hat{x})\delta\hat{\rho}\right) \label{eq:eom_2}
\end{align}
where we discarded second order terms in fluctuations. A useful tool to solve this linearized differential equation is the Laplace transform defined as
\begin{align}
    L[\hat{O}](s)=\int_{0}^\infty e^{-st}\hat{O}(t).
\end{align}

Applying the Laplace transform, we write out the two new equations
\begin{align}
sL[\delta\hat{\rho}]-\delta\hat{\rho}(0)=&\mathcal{L}_0L[\delta\hat{\rho}]-ig(L[\delta\alpha]+L[\delta\alpha^*])[\hat{\sigma}^x\cos(k\hat{x}),\hat{\rho}_0],\\
sL[\delta\alpha]-\delta\alpha(0)=&-(i\Delta+\kappa)L[\delta\alpha]-iNg\mathrm{Tr}\left(\hat{\sigma}^x\cos(k\hat{x})L[\delta\hat{\rho}]\right).
\end{align}

Solving for $L[\delta\hat{\rho}]$ in the top equation and substituting into the bottom equation we obtain
\begin{align}
A(s)L[\delta\alpha]+B(s)L[\delta\alpha^*]=\delta\alpha(0)-iNg\mathrm{Tr}\left(\hat{\sigma}^x\cos(k\hat{x})(s-\mathcal{L}_0)^{-1}\delta\hat{\rho}(0)\right)
\end{align}
with
\begin{align}
 A(s)=&s+\kappa+i\Delta+B(s),\\
 B(s)=&Ng^2\mathrm{Tr}\left(\hat{\sigma}^x\cos(k\hat{x})(s-\mathcal{L}_0)^{-1}[\hat{\sigma}^x\cos(k\hat{x}),\hat{\rho}_0]\right).
\end{align}
We now simplify the $B(s)$ term, which can be done by noting the relation $f(t) = e^{ht} \Rightarrow L[f](s) = \dfrac{1}{s-h}$. 

Therefore, applying the inverse Laplace transform we have
\begin{align}
\mathrm{Tr}\left(\hat{\sigma}^x\cos(k\hat{x})(s-\mathcal{L}_0)^{-1}[\hat{\sigma}^x\cos(k\hat{x}),\hat{\rho}_0]\right)
&= \int_{0}^{\infty}dt e^{-st}\mathrm{Tr}\left({\hat{\sigma}^x\cos(k\hat{x})e^{\mathcal{L}_0t}}[\hat{\sigma}^x\cos(k\hat{x}),\hat{\rho}_0]\right),\\
&= \int_{0}^{\infty}dt e^{-st}\left\langle[\hat{\sigma}^x(t)\cos(k\hat{x}(t)),\hat{\sigma}^x\cos(k\hat{x})]\right\rangle.
\end{align}
Here, we have used the adjoint channel of $\exp(\mathcal{L}_0t)$ to transform the initial time-independent operators $\hat{\sigma}_x$ and $\hat{x}$ into their time-dependent Heisenberg frame.
The time-evolution of $\hat{\sigma}^x$ and $\hat{x}$ governed by $\mathcal{L}_0$ is given by
\begin{align}
\hat{\sigma}^x(t)=&\hat{\sigma}^x\cos(\omega_0 t)-\hat{\sigma}^y\sin(\omega_0 t)\\
\hat{x}(t)=&\hat{x}+\frac{\hat{p}t}{m}.
\end{align}

We can split the expression into the two non-zero commutators
\begin{align}
   B(s)=& Ng^2\int_{0}^{\infty}dte^{-st}\left\langle[\hat{\sigma}^x(t)\cos(k\hat{x}(t)),\hat{\sigma}^x\cos(k\hat{x})]\right\rangle\\
   =&Ng^2\int_{0}^{\infty}dte^{-st}\left\langle\hat{\sigma}^x(t)\hat{\sigma}^x[\cos(k\hat{x}(t)),\cos(k\hat{x})]\right\rangle+Ng^2\int_{0}^{\infty}dte^{-st}\left\langle[\hat{\sigma}^x(t),\hat{\sigma}^x]\cos(k\hat{x})\cos(k\hat{x}(t))\right\rangle. \label{eq:B(s)_integral}
\end{align}
We now calculate the products inside the expectation values
\begin{align}
\langle\hat{\sigma}^x(t)\hat{\sigma}^x\rangle&=\langle\left(\hat{\sigma}^x\cos(\omega_0 t)-\hat{\sigma}^y\sin(\omega_0 t)\right)\hat{\sigma}^x\rangle \label{eq:spin_correlations1}\\
&=\cos(\omega_0 t)-i\sin(\omega_0 t)=e^{-i\omega_0 t},
\end{align}
where we explicitly used that the initial state is in the ground state and $\langle  {\downarrow}| {\hat{\sigma}^z}| 
 {\downarrow}\rangle=-1$.
For the terms including the position and momentum operators we use the Baker-Campbell-Hausdorff formula to evaluate 
\begin{align}
\cos(k\hat{x})\cos(k\hat{x}(t)) &= \dfrac{1}{2}\cos\left(\dfrac{kpt}{m}\right)e^{i\omega_Rt}.\label{eq:cos_cos_expansion}
\end{align}
Here we have introduced the recoil frequency $\omega_R=(\hbar k^2)/(2m)$. With this we can calculate
\begin{align}
    \langle\cos(k\hat{x})\cos(k\hat{x}(t))\rangle&=\frac{e^{i\omega_Rt}}{2}\int_{-\infty}^{\infty} dp \,\rho_p \cos\left(\frac{kpt}{m}\right) \\
&=\frac{e^{i\omega_Rt}}{2}\sqrt{\frac{\beta}{2\pi m}}\int_{-\infty}^\infty dp  \,e^{-\beta\frac{p^2}{2m}} \cos\left(\frac{kpt}{m}\right) \\
&=\frac{e^{i\omega_Rt}}{2}e^{-\frac{t^2}{2t^2_c}}\label{eq:MB_cos_cos_integral}
\end{align}
where we have used that the momentum distribution is a Maxwell-Boltzmann 
\begin{align}
    \rho_p=\sqrt{\dfrac{\beta}{2\pi m}}e^{-\beta\frac{p^2}{2m}},
    \end{align}
and introduced 
\begin{align}
  t_c=\sqrt{\frac{m\beta}{k^2}} \equiv \delta_D^{-1}.
\end{align}
Combining the results in Eqs.~\eqref{eq:spin_correlations1} and \eqref{eq:MB_cos_cos_integral} we can derive an expression for $B(s)$ that reads
\begin{align}
   B(s)
   =&-iNg^2\int_{0}^{\infty}dt e^{-st}e^{-\frac{t^2}{2t_c^2}}\sin([\omega_0+\omega_R]t).\label{eq:B_final_exp_spindown}
\end{align}

In order to find the dynamics of the field we need to solve
\begin{align}
    {\bf D}(s)\begin{pmatrix}
        L[\delta\alpha]\\
        L[\delta\alpha^*]
    \end{pmatrix}= \begin{pmatrix}
        \delta\alpha(0)-iNg\mathrm{Tr}\left(\hat{\sigma}^x\cos(k\hat{x})(s-\mathcal{L}_0)^{-1}\delta\hat{\rho}(0)\right)\\
        \delta\alpha^*(0)+iNg\mathrm{Tr}\left(\hat{\sigma}^x\cos(k\hat{x})(s-\mathcal{L}_0)^{-1}\delta\hat{\rho}(0)\right)
    \end{pmatrix}
\end{align}
with
\begin{align}
    {\bf D}(s)=\begin{pmatrix}
      A(s)&B(s)\\
      B^*(s)&A^*(s)
    \end{pmatrix}
    \label{eq:inst_matrix}.
\end{align}
Thus the Laplace transform of the field $L[\delta\alpha]$ can be found by inverting the matrix ${\bf D}(s)$. The dynamics of the real time $\delta\alpha$ is determined by the poles of ${\bf D}^{-1}(s)$ and therefore by the zeros of $D(s)=\det({\bf D}(s))$. To determine the stability of fluctuations we distinguish two scenarios: if all zeros of $D(s)$ have negative real part then the real-time evolution of $\delta\alpha$ will be damped and thus the initial configuration $\rho_0$ is stable. If at least one zero of $D(s)$ has positive real part then we expect exponential amplification and fluctuations will grow. In this case the initial state $\hat{\rho}_0$ is unstable. To find the threshold between the two we solve
\begin{align}
 D(0)=0.\label{eq:D00}
\end{align}

We use now Eq.~\eqref{eq:B_final_exp_spindown} and get
\begin{align}
D(0)=&\kappa^2+\Delta^2-2\Delta Ng^2\int_{0}^{\infty}dt e^{-\frac{t^2}{2t_c^2}}\sin([\omega_0+\omega_R]t).
\end{align}
Solving now Eq.~\eqref{eq:D00} for $g$ we get the critical coupling $g_c$ which is expressed as
\begin{align}
   \Omega_c=\frac{1}{\displaystyle\int_{0}^{\infty}dt e^{-\frac{t^2}{2t_c^2}}\sin([\omega_0+\omega_R]t)}.
\end{align}
This is the expression we outline in the main text, with $\Omega_c =-Ng_c^2\frac{2\Delta}{\kappa^2+\Delta^2} $. 

At last we want to discuss the steps to find the analytical results for the thermal $\delta_D\gg\omega_0+\omega_R$ and quantum regimes $\omega_0+\omega_R\gg\delta_D$.

\paragraph{Thermal regime}
For large temperatures (or the limit $\omega_0\rightarrow 0$), $[\omega_0+\omega_R]t_c\ll1$, we have
\begin{align}
    \int_{0}^{\infty}dt e^{-\frac{t^2}{2t_c^2}}\sin([\omega_0+\omega_R]t)=\int_{0}^{\infty}dt e^{-\frac{t^2}{2t_c^2}}[\omega_0+\omega_R]t=[\omega_0+\omega_R]t_c^2, 
\end{align}
and we get
\begin{align}
    Ng_c^2=\frac{\kappa^2+\Delta^2}{2\Delta}\frac{k^2}{m\beta[\omega_0+\omega_R]}=\frac{\kappa^2+\Delta^2}{2\Delta}\frac{2\omega_R}{\hbar\beta[\omega_0+\omega_R]}
\end{align}
which, for $\omega_0=0$, coincides with the classical threshold for self-organization \cite{Niedenzu:2011,Jaeger:2016}.
\begin{comment}
In the semiclassical analysis we are always in the regime where    $\omega_Rt_c\ll1\Leftrightarrow\sqrt{E_R\beta}\ll1
$
with $
    E_R=\hbar\omega_R.
$
Since we can still have $\omega_0 t_c>1$ for the semiclassical limit we should consider instead
\begin{align}
    \int_{0}^{\infty}dt e^{-\frac{t^2}{2t_c^2}}\sin([\omega_0+\omega_R]t)=\int_{0}^{\infty}dt e^{-\frac{t^2}{2t_c^2}}\left(\sin(\omega_0 t)+\cos(\omega_0 t)\omega_R t\right).
 \end{align}
\end{comment}

\paragraph{Quantum regime}
In the quantum regime, $\delta_D\rightarrow 0$, we have $(\omega_0+\omega_R) t_c\gg1$ which reduces our expression to
\begin{align}
    \int_{0}^{\infty}dt e^{-\frac{t^2}{2t_c^2}}\sin([\omega_0+\omega_R]t)\approx\frac{1}{\omega_0+\omega_R}.
\end{align}
Therefore, 
\begin{align}
    Ng_c^2=\frac{\kappa^2+\Delta^2}{2\Delta}\frac{k^2}{m\beta[\omega_0+\omega_R]}=\frac{\kappa^2+\Delta^2}{2\Delta}[\omega_0+\omega_R]
\end{align}
where we recover the phase transition of spinor self-ordering \cite{Mivehvar:2017,Kroeze:2018}. When additionally assuming $\omega_R\ll\omega_0$ we recover the standard result for the Dicke phase transition~\cite{Kirton:2019}.
\section{Numerical simulation of the dynamics}
In this section we provide additional details to the simulation method. 

The simulation method is a combination of discrete truncated Wigner simulation~\cite{Schachenmayer:2015b} for the atomic internal degrees of freedom and truncated Wigner for the atomic external degrees of freedom~\cite{Milburn:2008,Domokos:2001,Stenholm:1986}. The latter assumes that the atomic ensemble is in the semiclassical regime where the momentum recoil is very small compared to the single particle momentum width, $\omega_R\ll\Delta p$. This implies that $\delta_D\gg\omega_R$ which is outside of the validity regime of a generalized Gross Pitaevskii equation that are for instance used for spinor self-ordering~\cite{Mivehvar:2017}.  Our method enables the dynamical analysis of the system using trajectories of the semiclassical fields $a_x\sim (\hat{a}+\hat{a}^\dagger)/2$, $a_{p}\sim\mathrm{i}(\hat{a}^\dag-\hat{a})/2$ and positions and momenta $x_j$ and $p_j$. In addition, we sample semiclassical values for the Pauli matrices $s_j^\alpha\sim\hat{\sigma}_j^\alpha$ with $\alpha=x,y,z$. Their dynamics are coupled by the following stochastic differential equations
\begin{subequations}
\begin{align}
d_ta_x &= \Delta a_p  - \kappa a_x + \mathcal{N}_x \label{eq:cHL1}\\
d_ta_p &= -\Delta a_x - \kappa a_p  + NgX  + \mathcal{N}_p \\
d_ts^x_j &= -\omega_0 s^y_j\\
d_t s^y_j &= \omega_0 s^x_j - 2g\cos(kx_j)s^z_ja_x\\
d_t s^z_j &= -2g\cos(kx_j)s^y_ja_x\\
d_t x_j &= p_j/m\\
d_t p_j &= 2\hbar kg\sin(kx_j)s^x_ja_x. \label{eq:cHLend}
\end{align}    
\end{subequations}
We used $d_t=d/(dt)$, introduced the cavity shot noises $\mathcal{N}_x$ and $\mathcal{N}_p$ that are independent and defined by $\langle\mathcal{N}_x\rangle_{\mathrm{ens}}=0=\langle\mathcal{N}_p\rangle_{\mathrm{ens}}$ and $\langle\mathcal{N}_x(t)\mathcal{N}_x(t')\rangle_{\mathrm{ens}}=\kappa\delta(t-t')/2=\langle\mathcal{N}_p(t)\mathcal{N}_p(t')\rangle_{\mathrm{ens}}$, and $X\sim\hat{X}$ is the spin-self-organization order parameter in this description. We introduced here $\langle\,.\,\rangle_{\mathrm{ens}}$ which describes an average over several initializations. In order to describe the initial ground state we sample $s_j^\alpha=\pm1$ ($\alpha=x,y$) independently and $s_j^z=-1$. The atomic position are chosen randomly and uniformly from $[0,2\pi/k)$. This choice is justified since the above equations are periodic in space. The momenta are sampled from a Gaussian with zero mean and variance $\langle p^2\rangle_{\mathrm{ens}}=m/\beta$. To incorporate the initial vacuum fluctuations in the cavity field we sample $a_x$ and $a_p$ from a Gaussian with zero mean and $\langle a_x^2\rangle_{\mathrm{ens}}=1/4=\langle a_p^2\rangle_{\mathrm{ens}}$. We simulate this set of stochastic differential equations and the described initial conditions with a stability-optimized algorithm that makes explicit use of the fact that the noise terms are additive~\cite{Rackauckas:2017, Rackauckas:2020} The open-source implementation may be found in \cite{MN:github}.

\section{Experimental considerations}
 In this section, we connect the theoretical model studied in the main text to parameters attainable in the $^{87}\textrm{Rb}$ experiment in Tübingen~\cite{Suarez:2023}. With recoil frequency $\omega_R=\hbar k^2/2m=2\pi\times 38~\mathrm{kHz}$ the assumed thermal Doppler width of $\delta_D=10\,\omega_R$ corresponds to a temperature of $T=9~\mu\mathrm{K}$, which is within the reach of laser cooling. Also, by realizing the ground and excited state of the model Hamiltonian with Zeeman sublevels $m_F=\pm1$ of the $F=1$ - manifold of the $5S_{1/2}$ ground state, the transition frequency  can be tuned to $\omega_0=100\,\omega_R$ as in Fig.~2 in the main text by a moderate magnetic field of $B=1.1\,\mathrm{G}$. Moreover, a rescaled coupling strength of $\Omega=100\,\omega_R$ can be obtained with a number of $N=10\,000$ atoms, a usual laser intensity of $I=400\,\mathrm{mW/cm}^2$, an atomic detuning of $\Delta_a=2\pi\times1\,\mathrm{GHz}$, cavity detuning $\Delta=\kappa$ and a cavity with coupling constant $g_0=2\pi\times500\,\mathrm{kHz}$ and field decay rate $\kappa=2\pi\times300\,\mathrm{kHz}$. Thus, incoherent scattering is with $\gamma_\mathrm{scatt}=2\pi\times1.4\,\mathrm{kHz}$ negligible compared to all other rates. The corresponding light power leaking out of the cavity for typical values of $X^2=0.1$ reported in the main text is $P_\mathrm{out}=6.2\,\mathrm{nW}$ which can be used for detection of the phase transition. 
We thus conclude that we can implement the theoretical findings in our experiment.
\section{Determination of crossover from weak-to-strong spin self-organized phases}
In this section we will provide a formal definition of the \emph{persistence time}, $\tau$, that was used in the main text to distinguish between weak and strong spin self-organization. 

\begin{figure}[t!]
    \centering
    \includegraphics[width=0.75\textwidth]{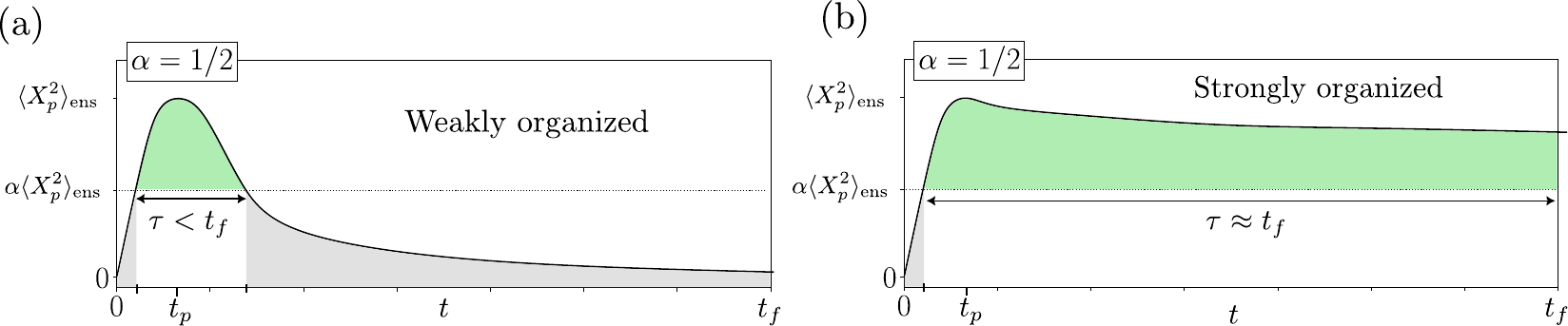}
    \caption{Schematic of the intuition behind the persistence time defined in \eqref{eq:persistence_time}. For the weakly organized phase, (a), once a peak value $\langle X_p^2\rangle_\text{ens}>0$  is reached, the system quickly relaxes back to $\langle X_f^2\rangle_\text{ens}\approx0$. Instead, the strongly organized regime, (b), is robust in time and there is negligible decay such that the final value at $t_f$ is close to the peak at $t_p$, $\langle X_f^2\rangle_\text{ens}\approx\langle X_p^2\rangle_\text{ens}$. With this in mind, $\tau$ gives us a time integrated measure of the robustness of the phase. Note the scale for $\langle X_p^2\rangle_\text{ens}$ is arbitrary and need not be equal for (a) and (b).}
    \label{fig:pers_time}
\end{figure}

\begin{figure}[h!]
    \centering
    \includegraphics[width=0.65\textwidth]{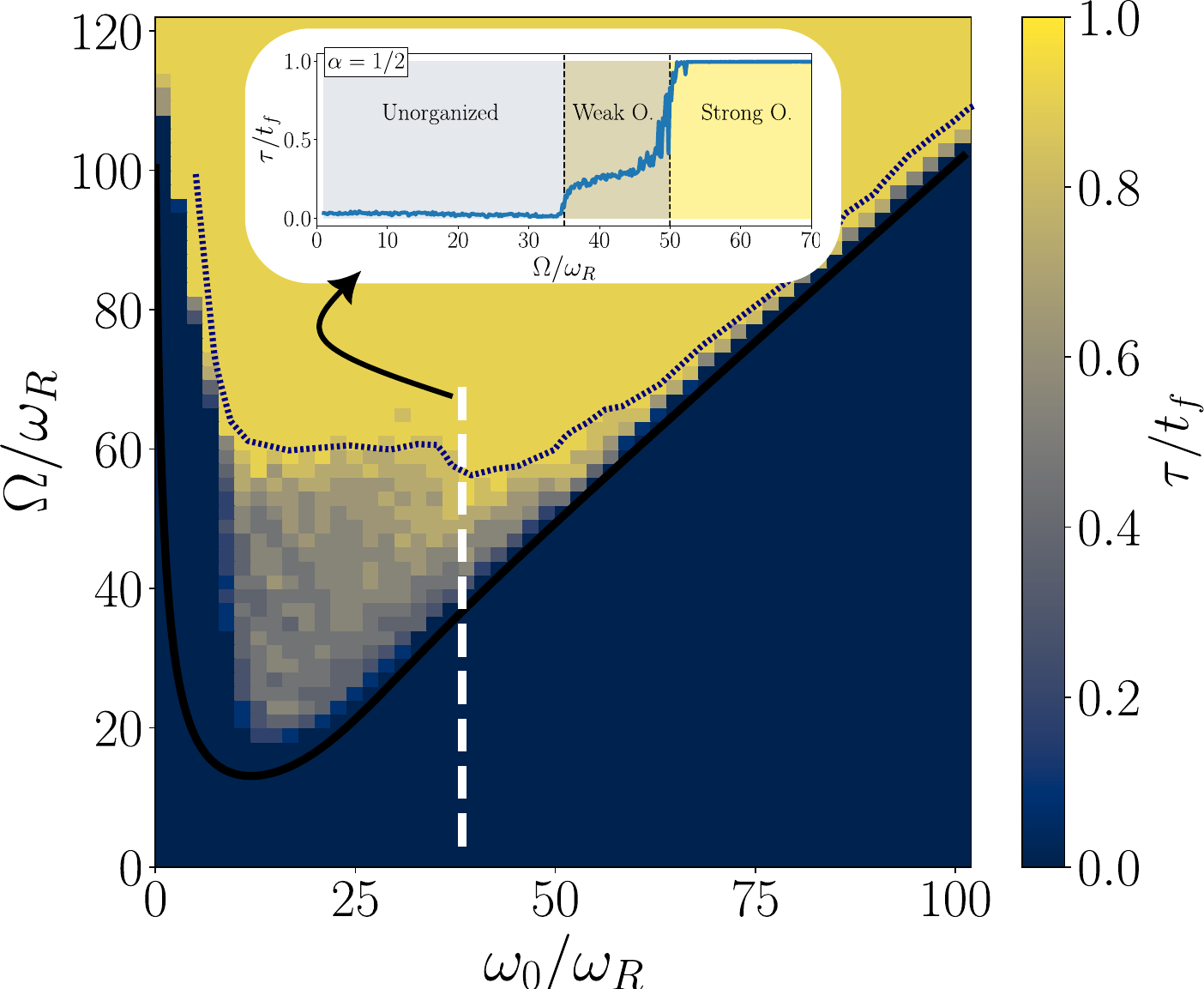}
    \caption{Phase diagram of the persistence time $\tau$, evaluated for $N=1600$ atoms, $\kappa=100\omega_R$, $\Delta=-\kappa$ and $\delta_D=10\omega_R$ with cavity shot-noise for simulation time $t_f\omega_R=20$. The lefternmost region appears more choppy as it is hard to distinguish the different regimes when the signal to noise ratio is low. The inset highlights how we can determine the transition between the transient weakly-organized to the robust strongly-organized phases. The crossover between the weak and strong organized phases (dotted blue line) is also plotted in Fig.~2c) and Fig.~2d) in the main text.}
    \label{fig:pers_time_phasediag}
\end{figure}

As discussed in the main text, a key result in our model is not only the existence of a spin self-organized phase but also a changing robustness of this phase depending on the location in parameter space. Inspired by similar survival analysis techniques \cite{kleinbaum:1996, bray:2013} we introduce the notion of \emph{persistence time}, $\tau$, to help us identify this behaviour. We define this as the fraction of evolution time the system remains above some threshold value $\alpha\langle X_p^2\rangle_\text{ens}$,  where $\alpha<1$ and $\langle X_p^2\rangle_\text{ens}$ is the peak value attained during the simulation (which will depend on the choice of parameters), see \sref{fig:pers_time} for a schematic. More formally, we define $\tau$ as the integral over the evolution time $T$
\begin{equation}
    \tau = \int_0^T\Theta\left(\langle X^2_p(t)\rangle_\text{ens}-\alpha\langle X^2_p\rangle_\text{ens}\right)\text{d}t \label{eq:persistence_time}
\end{equation}
 where $\Theta$ is the Heaviside step-function defined by $\Theta(x)=1$ if $x\geq0$ and $\Theta(x)=0$ otherwise. We can thus distinguish the different regimes in the spin self-organization phase diagram according to this parameter:\begin{equation}
 \begin{cases}
      \tau/T\simeq0, & \text{unorganized}, \\
      0<\tau/T<1, & \text{weakly organized}, \\
      \tau/T\simeq1, & \text{strongly organized}.
     \end{cases}
 \end{equation}
 The role of $\alpha$ is to smoothen or steepen the transition from the unorganized to the strongly organized phases depending if it is increased closer to 1 or reduced closer to 0, respectively. 
In our case, for the numerical analysis we have chosen $\alpha=1/2$ while changes of $\alpha$ result in slight modifications of the transition lines. We plot the phase diagram of the persistence time $\tau$ for $\alpha=1/2$ in Fig.~S\ref{fig:pers_time_phasediag}. The dashed line that marks the transition from weak to strong spin self-organization is also visible in Fig. 2 in the main text. 

\section{Finite size scaling}
In this section we discuss the finite size scaling of the quantities that are reported in the main text.

\subsection{Spin self-organization}
In Fig.~S\ref{fig:Cos2adaga_finsize} we present the ensemble averaged values of $\mathcal{B}$ in (a) and (b) and of $X^2$ in (c) and (d) for different atom numbers $N=100,$ $800$, $1600$ over a timescale of $t_f=20/\omega_R$. While the changes from $N=100$ to $N=800$ are still significant we only observe minor changes wen changing from $N=800$ to $N=1600$. We believe that the obtained results are therefore close to the ones expected for $N\to\infty$.
\begin{figure}[h!]
\includegraphics[width=0.49\linewidth]{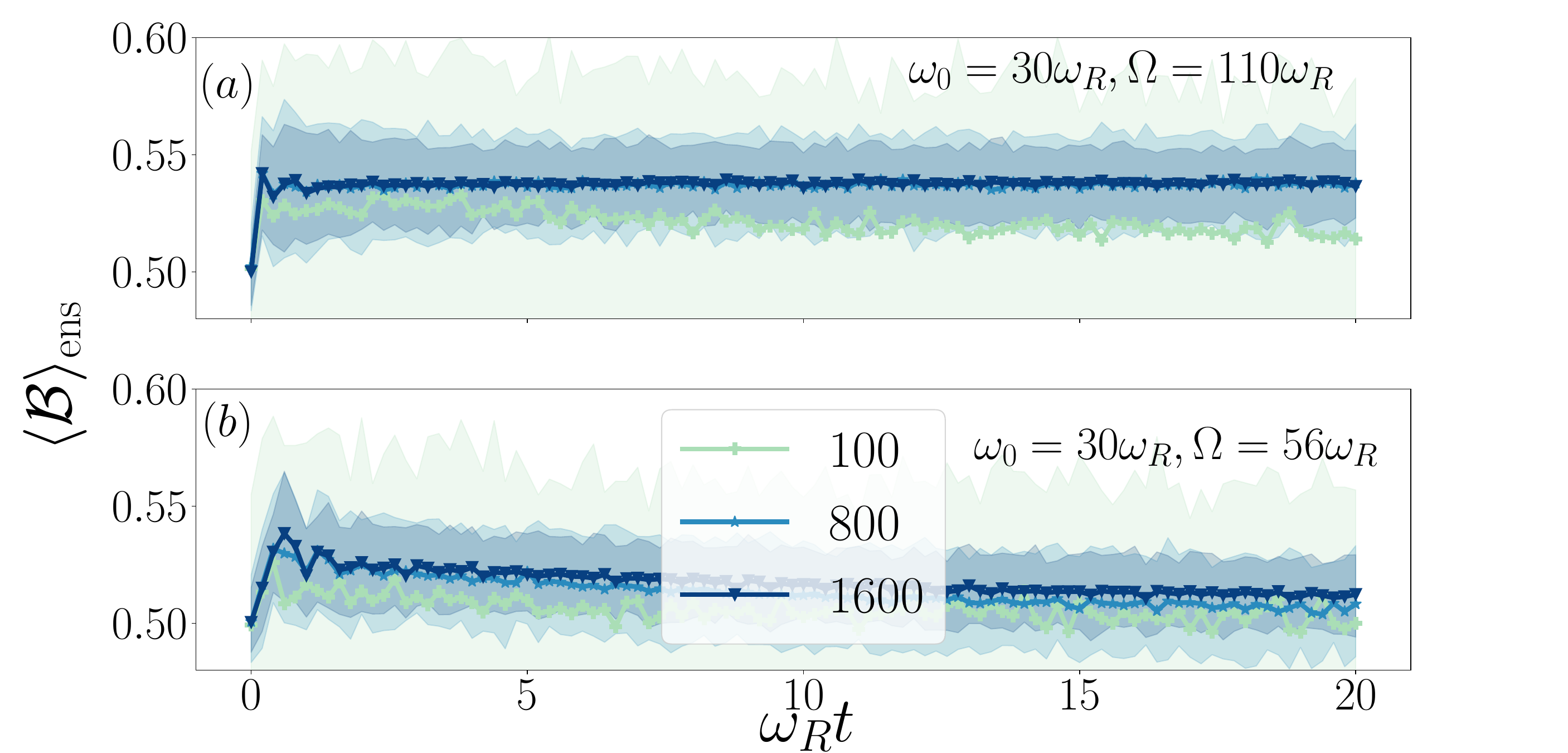}
 %   \centering
 %   \begin{subfigure}[b]{0.50\linewidth}
 %       \centering
 %       \includegraphics[width=1.\linewidth]{figures/bunch_finsize_stoch.pdf}
 %   \end{subfigure}%
 %   \hfill
 %   \begin{subfigure}[b]{0.50\linewidth}
  %      \centering
        \includegraphics[width=0.49\linewidth]{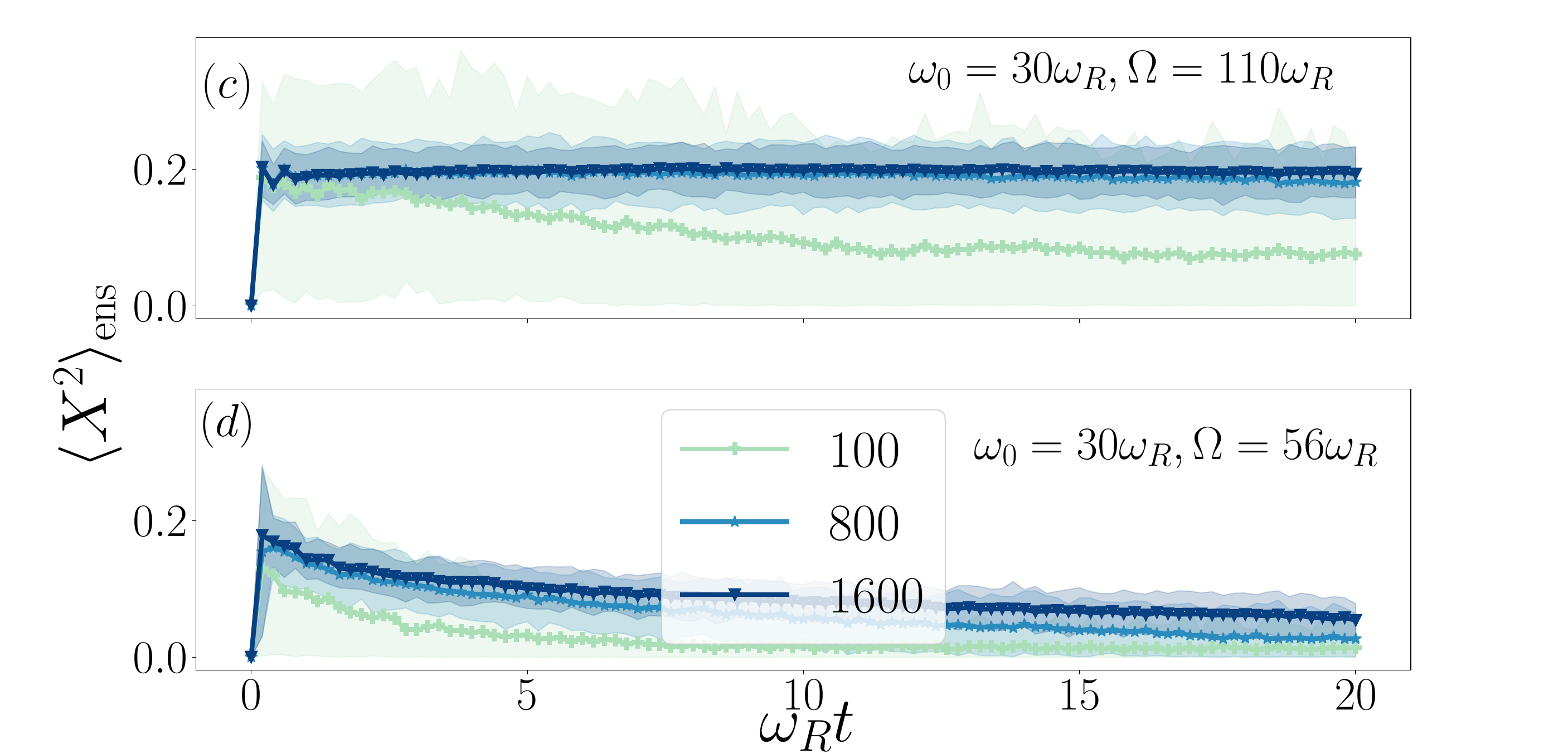}
  %  \end{subfigure}
    \caption{Finite size scaling of bunching (a,b) and spin self-organization order parameter (c,d) for 100 trajectories for different system sizes. (a) and (c) correspond to the strongly organized regime, (b) and (d) one to the weakly organized one. Solid markers represent the mean and shaded regions bound 90\% of trajectories.}
    \label{fig:Cos2adaga_finsize}
\end{figure} 
In the strong spin self-organized regime (a) and (c) we also find almost constant values for the ensemble averaged quantities on the studied timescales. For the weak spin self-organized regime represented by (b) and (d) we see decaying dynamic of the bunching and order parameters. This is in agreement with the claims in the main text that the pattern decay over long timescales and dicriminates the regimes of strong and weak spin self-organized regimes. We remark the in the weak spin self-organized regimes it seems that the dynamics of the bunching parameter is slightly lower than the one of the order parameter.

\subsection{Momentum statistics}
We now study the dynamics of quantities that describe the momentum statistics of the atoms. We use the same color coding and the same parameters as in the previous subsection. In (a) and (b) we show the mean kinetic energy and in (c) and (d) the kurtosis which was introduced in the main text. We find that all curves have converged for $N=1600$ and therefore conclude that the results are close to the ones expected in the thermodynamic limit $N\to\infty$.
    \begin{figure}[h!]
%    \centering
%    \begin{subfigure}[b]{0.50\linewidth}
%        \centering
        \includegraphics[width=0.49\linewidth]{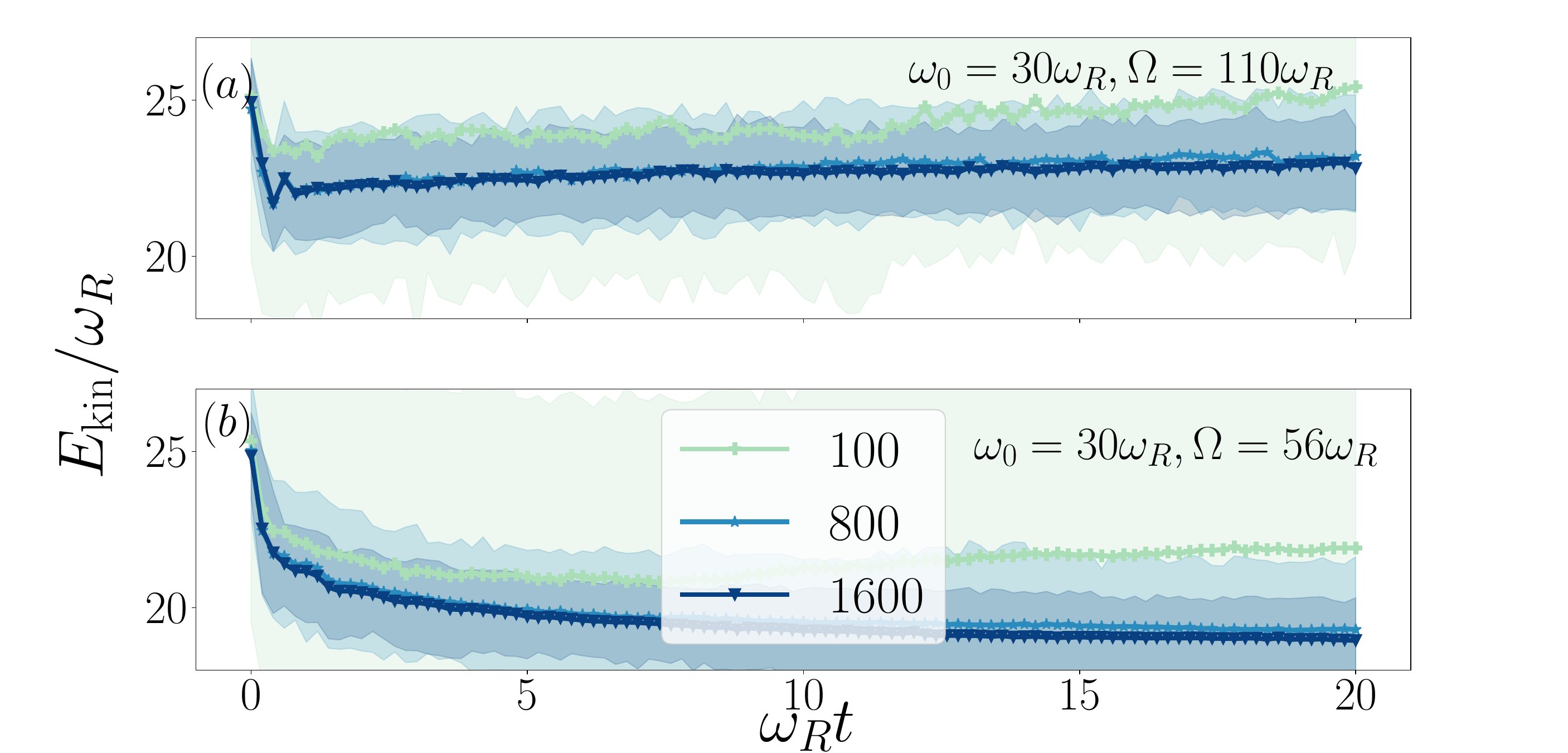}
    %\end{subfigure}%
    %\hfill
    %\begin{subfigure}[b]{0.50\linewidth}
     %   \centering
        \includegraphics[width=0.49\linewidth]{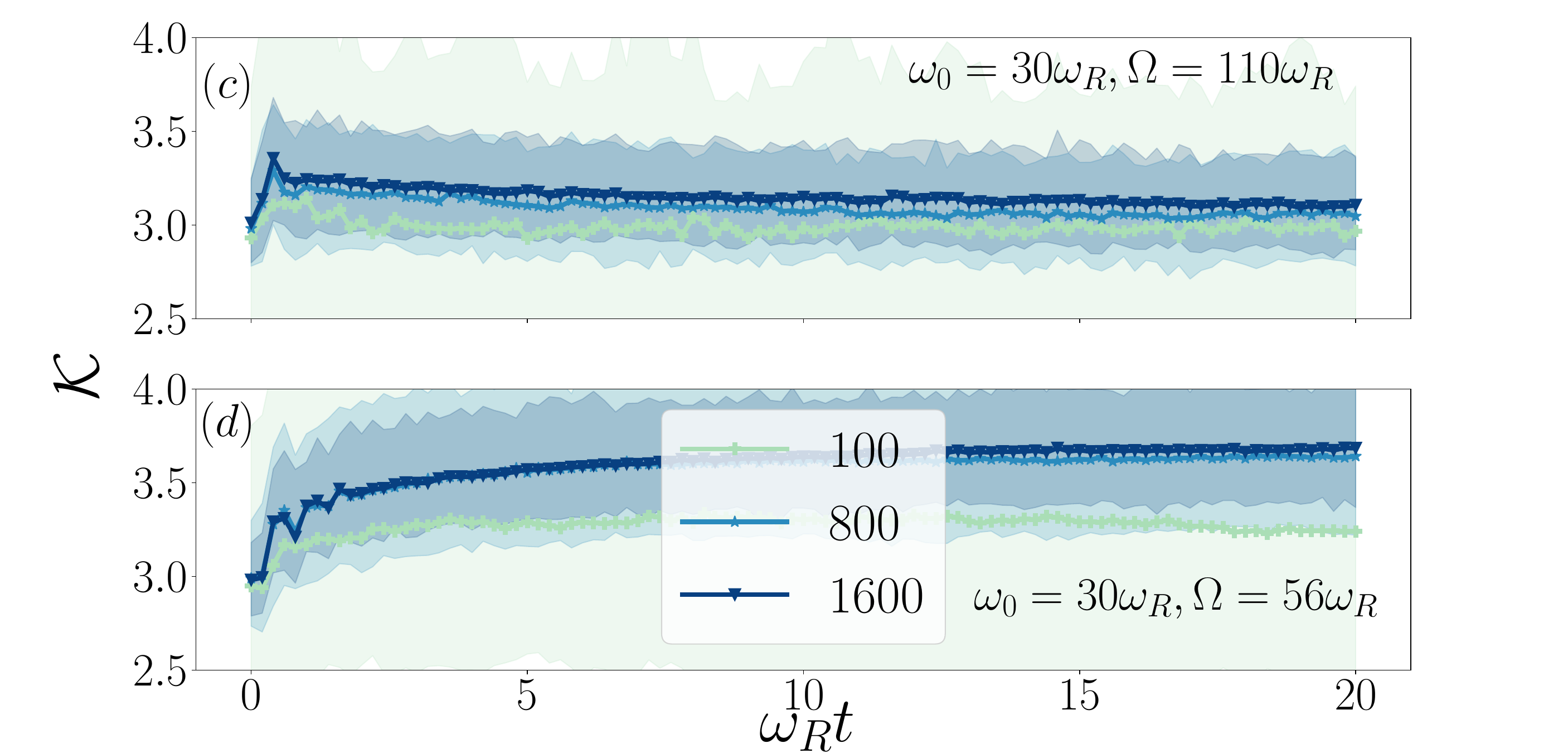}
    %\end{subfigure}
    \caption{Finite size scaling of kinetic energy (a,b) and kurtosis (c,d) for 100 trajectories for different system sizes. (a) and (c) correspond to the strongly organized regime, (b) and (d) one to the weakly organized one. Solid markers represent the mean and shaded regions bound 90\% of trajectories.}
    \label{fig:Cos2adaga_finsize}
\end{figure} 
The time evolution of the kinetic energy and kurtosis in the strong spin self-organized regime visible in (a) and (c) only show a small deviation from their initial values. In contrast, in the weak spin self-organized regime visible in (b) and (d), we find a tremendous reduction of the kinetic energy and a significant increase of the kurtosis. These findings are discussed in the main text.

\section{Other parameter choices}
In this section we report simulations for different ratios of $\delta_D/\omega_R$. The purpose of this section is to demonstrate the general statements made in the main text do not depend on the ratio of $\delta_D/\omega_R$ but on the ratio of $\delta_D/\omega_0$ and the coupling strength $\Omega$.

\subsection{Spin self-organization}
\begin{figure}[h!]
    \centering
    \includegraphics[width=0.85\textwidth]{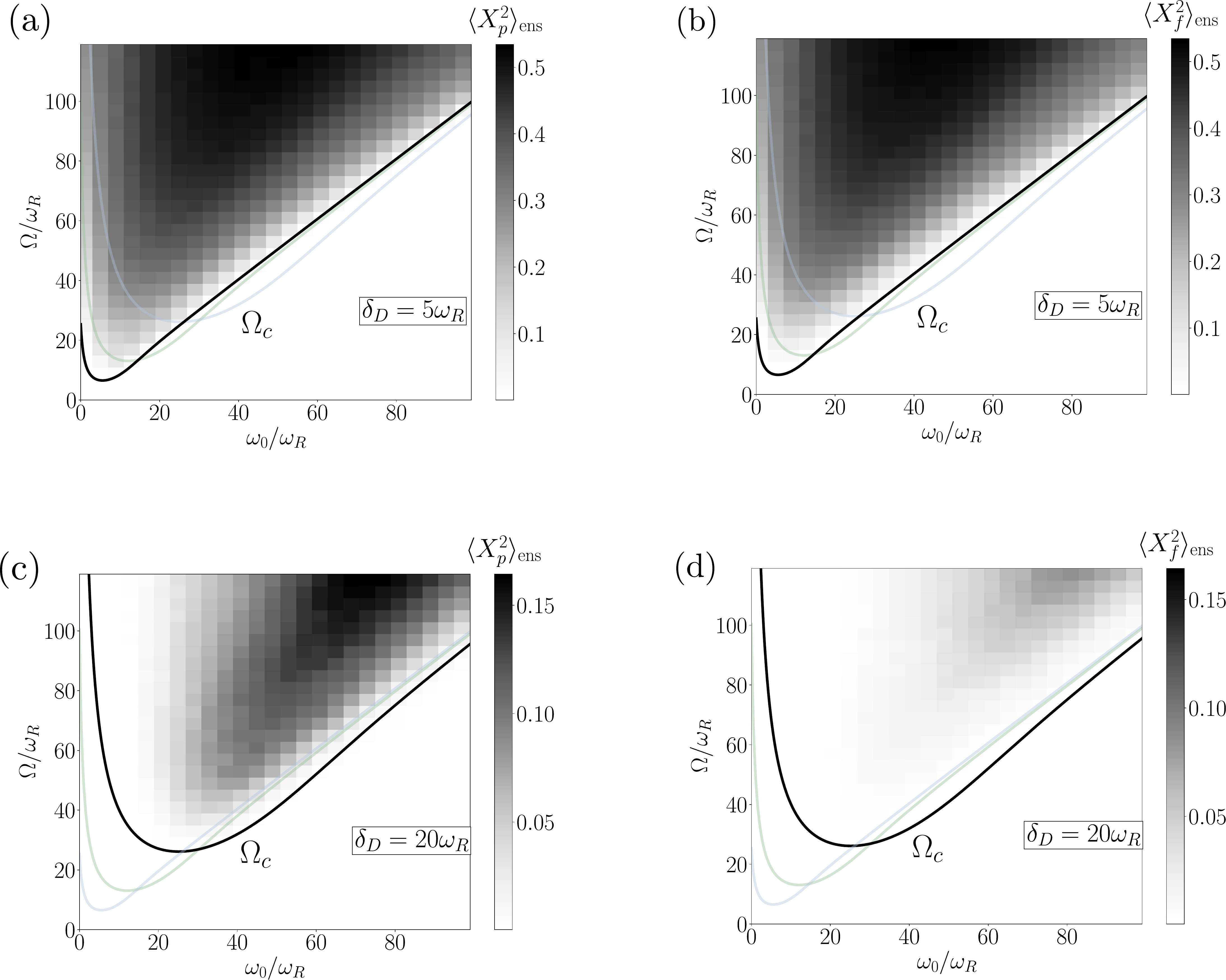}
    \caption{Phase diagrams of the spin self-organized order parameter, $\langle X^2\rangle_\text{ens}$ for Doppler widths $\delta_D=5\omega_R$ (a,b) and $\delta_D=20\omega_R$ (c,d), in the peak (a,c) and final (b,d) regimes. The solid black line indicates the analytical threshold $\Omega_c$ for the $\delta_D$ of choice. The differently colored shaded lines correspond to the thresholds of the other Doppler widths studied. One can spot the linear scaling of the threshold minima with a changing $\delta_D$.}
    \label{fig:phase_diagrams_deltas}
\end{figure}

\begin{figure}[h!]
    \centering
    \includegraphics[width=0.85\textwidth]{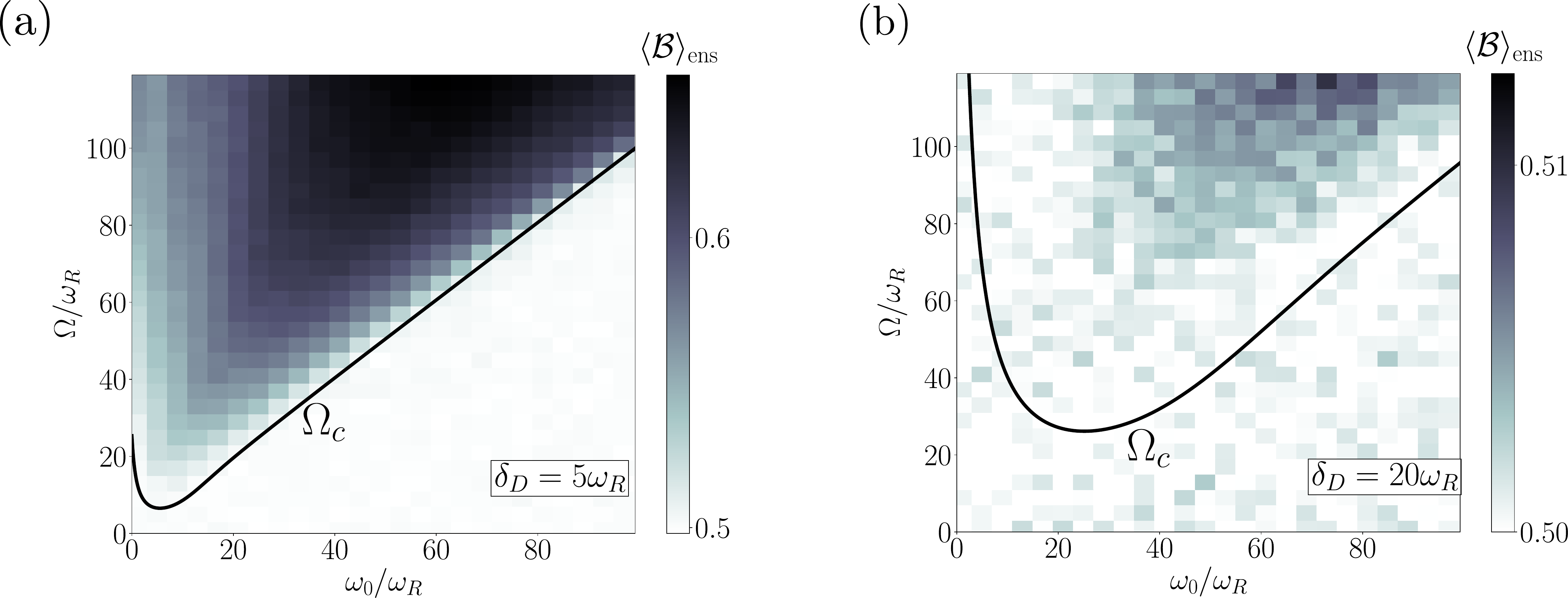}
    \caption{Phase diagrams of the final bunching parameter, $\langle\mathcal{B}\rangle_\text{ens}$ for Doppler widths $\delta_D=5\omega_R$ (a) and $\delta_D=20\omega_R$ (b).}
    \label{fig:phase_diagrams_deltasCos2}
\end{figure}

In Fig.~S\ref{fig:phase_diagrams_deltas} we show the diagrams for the $\langle X^2(t)\rangle_{\mathrm{ens}}$ at times $t=t_p$in (a) and (c) and $t=t_f$ in (b) and (d) and for ratios $\delta_D/\omega_R=5$ in (a) and (b) and $\delta_D/\omega_R=20$  in (c) and (d).
 We remark that for the corresponding choices the minimum threshold is shifted to a value with matches $\omega_0\sim\delta_D$. Comparing the subfigures we find that the parameter regime where weak spin self-organization is found is much larger for larger Doppler width $\delta_D=20\omega_R$. This is consistent with our claim that weak spin self-organization requires $\delta_D\sim\omega_0$ and $\Omega<\Omega_c^{\mathrm{Self-org.}}$. The threshold value $\Omega_c^{\mathrm{Self-org.}}=25\omega_R(400\omega_R)$ for the values of $\delta_D=5\omega_R(20\omega_R)$. Thus most parameters visible in Fig.~S\ref{fig:phase_diagrams_deltas} (a) and (b) fulfill already $\Omega>\Omega_c^{\mathrm{Self-org.}}$ while every parameter in (c) and (d) fulfills $\Omega<\Omega_c^{\mathrm{Self-org.}}$. This explains why weak spin self-organization appears to be the most common feature for $\delta_D=20\omega_R$ while it only appears on the very minimum of the threshold for $\delta_D=5\omega_R$. This is also visible in the bunching parameter which is shown in Fig.~S\ref{fig:phase_diagrams_deltasCos2}. 

\subsection{Momentum statistics}
\begin{figure}[h!]
    \centering
    \includegraphics[width=0.85\textwidth]{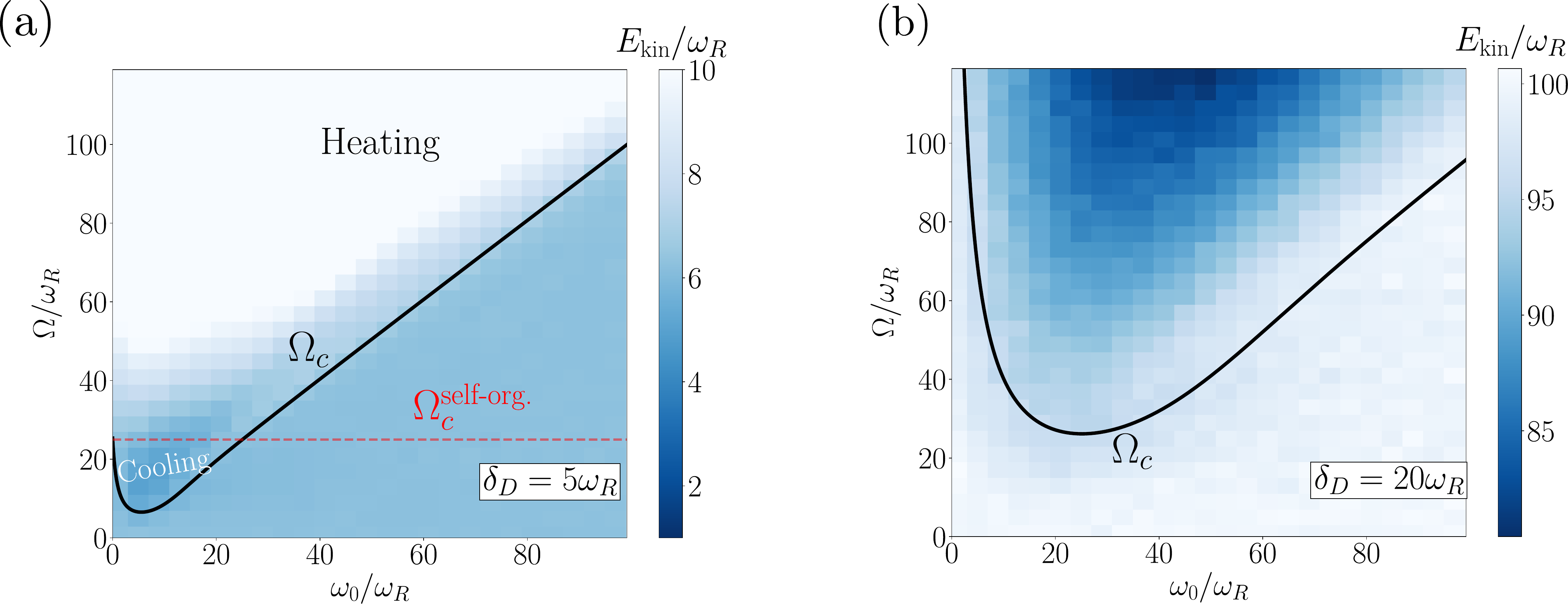}
    \caption{Phase diagrams of the final kinetic energy, $E_\text{kin}$ for Doppler widths $\delta_D=5\omega_R$ (a) and $\delta_D=20\omega_R$ (b). In the case of $\delta_D=5\omega_R$ we can spot regions of both cooling and heating, only present below or above the classical $\Omega_c^\text{self-org.}$ line respectively. Note the different energy scales. }
    \label{fig:phase_diagrams_deltasEkin}
\end{figure}

\begin{figure}[t!]
    \centering
    \includegraphics[width=0.85\textwidth]{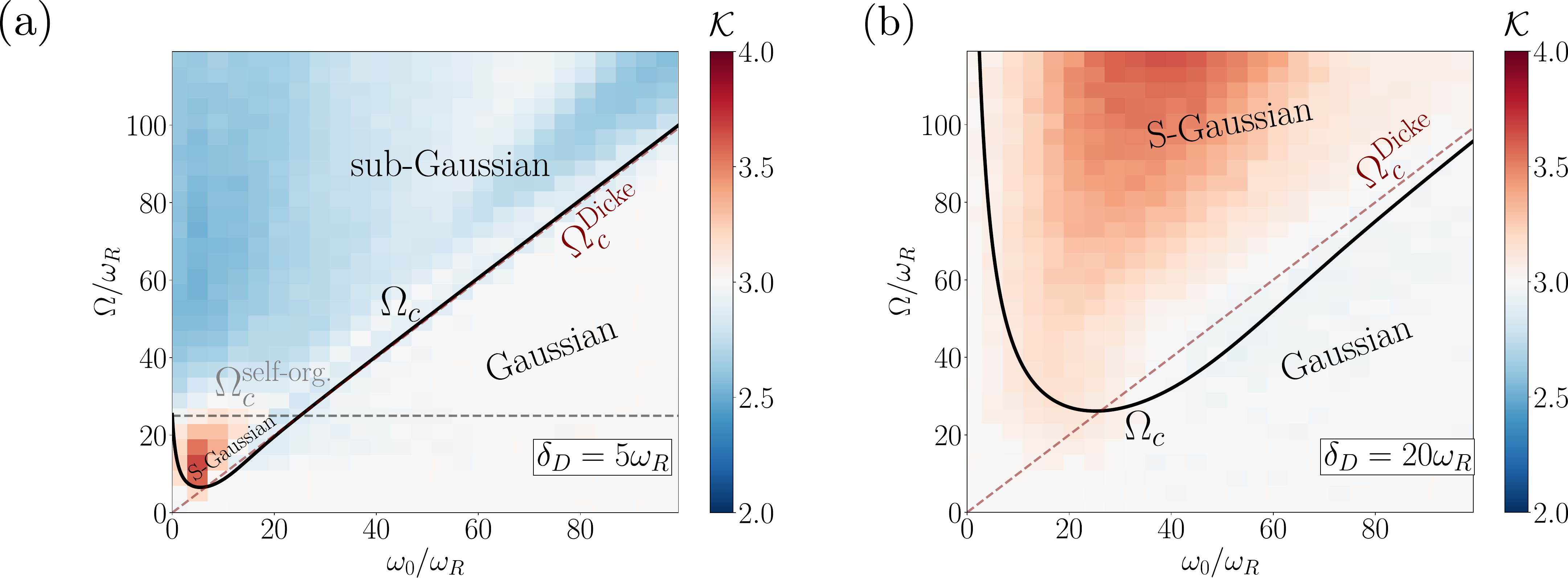}
    \caption{Phase diagrams of the final kurtosis, $\mathcal{K}$ for Doppler widths $\delta_D=5\omega_R$ (a) and $\delta_D=20\omega_R$ (b). Note the reduced Super-gaussian regime in the $\delta_D=5\omega_R$ case is fully enclosed by the area between the classical self-organization and the quantum Dicke threshold. The new region of sub-Gaussianity is present only above the classical $\Omega_c^\text{self-org.}$ threshold. In the $\delta_D=20\omega_R$ case the classical self-organization line lies well above the edge of the diagram at $\Omega_c^\text{self.org}=400\omega_R$.}
    \label{fig:phase_diagrams_deltaskurt}
\end{figure}

We now study the momentum statistics for the same parameters that have been investigated in the previous subsection. 
In Fig.~S\ref{fig:phase_diagrams_deltasEkin}
 we show the mean kinetic energy for $\delta_D=5\omega_R$ and $\delta_D=20\omega_R$ in (a) and (b). We find that a reduction of kinetic energy is visible almost everywhere for $\delta_D=20\omega_R$ in subplot (b) but only in a very small parameter regime for $\delta_D=5\omega_R$ in (a). We find that this is mostly due to the value of $\Omega_c^{\mathrm{Self-org.}}=25\omega_R$ for $\delta_D=5\omega_R$ [see red dashed line in (a)]. 
Even more striking is the difference in the kurtosis that is visible in Fig.~S\ref{fig:phase_diagrams_deltaskurt}. In (a) for $\delta_D=5\omega_R$ we find $\mathcal{K}>3$ only for the very small parameter regime set by $\Omega_c^{\mathrm{Self-org.}}=25\omega_R$. In (b) we find $\mathcal{K}>3$ everywhere in the spin self-organized regime due to the fact that $\Omega_c^{\mathrm{Self-org.}}=400\omega_R$ which is larger than the simulated range of $\Omega$.

The supplemental information about the kinetic energy and kurtosis for very different parameters highlight the generality of our results. The features discussed in the main text that result in a reduction of the kinetic energy and an enhanced kurtosis mostly depend on  $\omega_0\sim\delta_D$ and $\Omega<\Omega_c^{\mathrm{Self-org.}}.$